\documentclass[floatfix,prd,lengthcheck,showpacs,amssymb,amsmath,amsfonts,aps,altaffilletter,nofootinbib,nopreprintnumbers,showpacsm]{revtex4-1}

 \usepackage[utf8]{inputenc}
\usepackage{color}
\usepackage{graphicx}
\usepackage{tabularx}
\usepackage{float}
\usepackage{subfigure}
\usepackage{lipsum}  
\usepackage{amsmath}
\usepackage{acronym}
\usepackage{multirow}
\usepackage{tabu}
\usepackage{tikz}
\usetikzlibrary{arrows,shapes,trees,decorations.pathreplacing}
\usepackage{import}
\usepackage{svn-multi}
\usepackage{hyperref}
\usepackage{xspace}
\hypersetup{
    colorlinks=true,
    linkcolor=blue,
    filecolor=magenta,      
    urlcolor=cyan,
}

\newcommand{\mnras}{MNRAS}

\begin{document}

\title[]{Does non-stationary noise in LIGO and Virgo affect the estimation of $H_0$?}
\author{Simone Mozzon}
\email{simone.mozzon@port.ac.uk} 
\author{Gregory Ashton}
\author{Laura K. Nuttall}
\author{Andrew R. Williamson}
\affiliation{University of Portsmouth, Portsmouth, PO1 3FX, UK}

\date{\today}

\begin{abstract}
Gravitational-wave observations of binary neutron star mergers and their electromagnetic counterparts provide an independent measurement of the Hubble constant, $H_0$, through the standard-sirens approach. Current methods of determining $H_0$, such as measurements from the early universe and the local distance ladder, are in tension with one another.
If gravitational waves are to break this tension a thorough understanding of systematic uncertainties of gravitational-wave observations is required. To accurately estimate the properties of gravitational-wave signals measured by LIGO and Virgo, we need to understand the characteristics of the detectors noise. Non-gaussian transients in the detector data and rapid changes in the instrument, known as non-stationary noise, can both add a systematic uncertainty to inferred results. We investigate how non-stationary noise affects the estimation of the luminosity distance of the source, and therefore of $H_0$.
Using a population of 100 simulated binary neutron star signals, we show that non-stationary noise can bias the estimation of the luminosity distance by up to 6.8\%.
However, only $\sim$15\% of binary neutron star signals would be affected around their merger time with non-stationary noise at a similar level to that seen in the first half of LIGO-Virgo's third observing run. Comparing the expected bias to other systematic uncertainties, we argue that non-stationary noise in the current generation of detectors will not be a limiting factor in resolving the tension on $H_0$ using standard sirens. Although, evaluating non-stationarity in gravitational-wave data will be crucial to obtain accurate estimates of $H_0$.
\end{abstract}

\maketitle

\section{Introduction}\label{s:intro}

Observations of binary neutron star (BNS) mergers and their electromagnetic (EM) counterparts can probe the expansion history of the universe \cite{Schutz:1986ss,Holz_2005}. Gravitational waves from compact binary coalescences (CBC) are standard sirens, which means that identical mergers always have the same luminosity. 
Therefore, we can directly estimate the luminosity distance of their source. 
EM counterparts, such as kilonovae or gamma-ray bursts, allow astronomers to identify the host galaxies \cite{Dalal_2006, Nissanke_2010} from which we can measure the cosmological redshift. The relationship between the luminosity distance of the source and its redshift depends on cosmological parameters, and, for late times, is dominated by the Hubble constant ($H_0$).

Currently, measurements of the Cosmic Microwave Background \cite{refId0} are in tension with observations based on the local distance ladder \cite{Riess_2019}; this tension has risen to the $4.4 \sigma$ level \cite{Riess_2019,DiValentino:2021izs,Freedman:2021ahq}.
Being completely independent and a self-calibrated measurement of $H_0$, standard sirens could have a crucial role in solving this tension. The first estimation of $H_0$ from standard sirens \cite{LIGOScientific:2017adf} gave results broadly consistent with other measurements found to date \cite{freedman2017cosmology}. Multiple multimessenger detections (i.e. gravitational-waves and EM detections) are required to improve the accuracy; several studies predict a $1\%$ $H_0$ measurement accuracy is achievable with $\mathcal{O}(100)$ detections~\cite{nissanke2013determining, Chen_2018, Feeney_2019, 2019PhRvD.100j3523M}. This would break the tension on $H_0$. 

An accurate estimation of $H_0$ will depend crucially on the understanding of the systematic uncertainties in both EM and gravitational-wave observations. One source of systematic error related to the observation of the EM counterparts regards the peculiar velocity field of the host galaxy \cite{10.1093/mnras/staa049,10.1093/mnras/staa1120,Mukherjee_2021}. The uncertainty on the peculiar velocity is dominant only for extremely close events and is negligible for most of the expected future detections. An additional bias can arise from mis-modelling the kilonova signal on the inclination \cite{Chen_2020}. Improved models of the kilonova emission could reduce this uncertainty, however, this will require significant theoretical progress \cite{10.1093/mnras/stab221}. 
Currently, the known dominant systematic uncertainty in the standard sirens approach is due to the gravitational-wave data. The main source is the detectors calibration, with an uncertainty in the amplitude of the calibrated strain below $2\%$ in both LIGO~\cite{Sun_2020, 2015CQGra..32g4001L} and Virgo~\cite{TheVirgo:2014hva, Estevez_2021} detectors. This uncertainty should decrease in future observing runs, but, even at the current level, does not limit the resolution of the $H_0$ tension \cite{arxiv.2204.03614}. 
An unaccounted source of systematic uncertainty could arise from mis-modelling the noise in the gravitational-wave detectors in the estimation of the luminosity distance. Indeed, the most widely used inference codes for gravitational waves -- (\texttt{LALInference} \cite{Veitch_2015}, \texttt{Bilby} \cite{Ashton_2019,10.1093/mnras/staa2850}, \texttt{PyCBC Inference} \cite{Biwer_2019} and \texttt{RIFT} \cite{lange2018rapid}) -- assume that the detector noise is both stationary and Gaussian \cite{PhysRevD.84.122004}. Gaussianity refers to the distribution of the noise and means that the noise can be completely characterised by a mean vector and a covariance matrix. Stationary noise means that the statistical properties of the noise do not vary in time.

In reality, due to broadband sources of noise of instrumental or environmental origin, data from ground-based detectors, such as LIGO and Virgo, are both non-Gaussian and non stationary \cite{Abbott_2016,Abbott_2020,RICHABBOTT2021100658,Davis_2021}. Non-Gaussianities are generally noise transients (called glitches \cite{Nuttall_2015, Zevin_2017}) that last on the order of a second. Non-stationary noise, instead, can vary the detector sensitivity on the order of tens of seconds affecting especially long duration signals. Noise transients are more obvious within gravitational-wave data compared to non-stationary noise. As such, analyses can be performed with noise transients either subtracted or the frequency range of analyses restricted to limit the impact of the noise transient (e.g.~\cite{LIGOScientific:2017vwq, Pankow_2018, LIGOScientific:2020ibl}). It is not possible to employ these workaround techniques with non-stationary noise as the noise tends to be more subtle, harder to model and the noise usually impacts a large frequency range.


The effect of mis-modelling the noise in the parameter estimation of gravitational-wave signals can be estimated analytically by assuming the \textit{linearised signal approximation} (LSA) \cite{PhysRevD.77.042001}, where the
template waveform $h(\theta)$ is expanded as a linear function of the true signal $h_0$ across the expected uncertainties of the parameters ~\cite{Edy_2021}. With this approximation and using an uninformative prior, the maximum likelihood of the parameters averaged over non-stationary Gaussian noise realisations is an unbiased estimator of the true source parameters, which means that mis-modelling the noise does not affect the posterior mode. Non-stationarity affects only the uncertainty on the posteriors, which is mis-estimated in particular for longer signals like BNS ~\cite{Edy_2021}. 

Although the LSA is a good baseline to understand the effect of non-stationary noise in simple cases, it is impractical for real data, being valid just for high SNR signals. As shown in ~\cite{PhysRevD.77.042001}, to correctly estimate the parameters for low-SNR signals requires higher orders in the template expansion. 
Moreover, the prior can not be easily handled analytically for non-flat or non-Gaussian priors. Including an informative prior can bias the estimation, introducing noise dependent terms in the posterior mode. As discussed in section \ref{Section:Injection}, the luminosity distance is generally estimated by adopting a uniform prior in Euclidean volume, therefore mis-estimating the noise could bias the luminosity distance posterior.

In conclusion, non-stationary noise could affect the estimation of the luminosity distance, and, consequently, the number of detections necessary to reach a few percent measurement of $H_0$. More importantly, non-stationary noise could bias the estimation of $H_0$. 

In this paper we investigate how non-stationary noise affects the estimation of the luminosity distance of BNS signals from gravitational-wave data, assuming the detection of an electromagnetic counterpart. We use publicly available LIGO and Virgo data from the first half of the third observing run (O3a) \cite{Vallisneri_2015, RICHABBOTT2021100658}.
While previous studies have investigated how to obtain more accurate parameter estimation in non-stationarity data \cite{PhysRevD.102.124038, PhysRevD.103.044006}, fully accounting for non-stationarity in parameter estimation is still computationally prohibitive. Therefore, we aim to determine if this effort is necessary to solve the Hubble tension. 

The rest of the paper is organised as follows. Section \ref{Section:Basics non-stat} describes 
the Bayesian approach which is widely used to estimate the parameters of gravitational-wave signals and how non-stationarity breaks its basic assumptions. In section \ref{Section:Method} we introduce our investigation on the effect of non-stationary noise in the estimation of the luminosity distance. We also present our results discussing the possible consequences on the estimation of $H_0$ through gravitational-wave data. In section \ref{Section:Conclusions} we summarise our main results and conclude.

\section{Parameter estimation in non-stationary noise}\label{Section:Basics non-stat}

The output of a gravitational-wave interferometer is a time series $d(t)$ such that:
\begin{equation}
    d(t)= \begin{cases}
        n(t)+h(t), & \text{if a signal is present,}\\
        n(t), & \text{otherwise,}
        \end{cases}
\end{equation}
where $n(t)$ is the detector noise and $h(t)$ is a gravitational-wave signal. Gravitational-wave transient searches identify signals by matched-filtering the data with a number of templates which sample the waveform parameter space \cite{Cutler_1993, Allen_2012}. Once the merger time is identified, the posterior probability densities for the source parameters are extracted using a Bayesian approach \cite{PhysRevD.46.5236, Romano_2017}. This approach requires the computation of the likelihood function, which represents the probability of observing data $d$ assuming that the signal has parameters $\theta$. If the noise is Gaussian with zero mean, the single detector likelihood takes the form \cite{Veitch_2010, LIGOScientific:2017vwq}
\begin{equation}\label{like}
    \mathcal{L} = p(d | h(\theta)) = \frac{1}{det(2\pi C_n)}e^{-\frac{1}{2} r^{\dagger}(f)C_n^{-1}r(f)}
\end{equation}
where the residual $r(f) = d(f)-h(f,\theta)$ is assumed to have the same distribution as the noise. $C_n=\langle n^{*}(f) n(f') \rangle$ is the noise covariance matrix, where the angle brackets denote averaging over different realisations of the noise. We use the variables $t$ and $f$ to indicate whether a quantity is in the time or frequency domain. 
If the data are stationary the frequencies are completely uncorrelated. In this case, ${\langle n^{*}(f) n(f') \rangle}$ is diagonal and is fully described by the noise power spectral density (PSD) $S_n(f)$:
\begin{eqnarray}\label{PSD}
\langle n^{*}(f) n(f') \rangle = \frac{T}{2} S_n (|f|) \delta(f - f') ~,
\end{eqnarray}
where T is the duration of the analysed data. Combining Equations~\eqref{like} and \eqref{PSD} we obtain the likelihood function typically used for gravitational-wave parameter estimation \cite{Veitch_2015}. This model is accurate for short segments of data from ground based interferometers. However, the assumption of stationary breaks down for periods of 64 seconds \cite{PhysRevD.91.084034}, a smaller window than what is typically needed to analyse binary neutron star mergers. Non-stationarity appears in Equation \eqref{like} as off-diagonal terms in the covariance matrix \cite{talbot2021inference}. Ref. \cite{Edy_2021} showed that ignoring these terms would affect the width of the posterior, with more evident effects when the signal is longer in duration. Nevertheless, accounting for non-stationarity would increase the computational cost from $\mathcal{O}(N)$ to $\mathcal{O}(N^2)$, where $N^2$ is the number of elements in the covariance matrix. This would be prohibitive in particular for longer signals.

Even assuming a diagonal covariance matrix as a fair approximation, non-stationary noise would bias the estimation of the noise spectrum. As a workaround, a common approach is to compute the PSD ``off-source'', using data close to, but not containing, the detected signal. However, this approach has an intrinsic uncertainty which could introduce new biases in the parameter estimation \cite{2020PhRvR...2d3298T}. Moreover it produces poor estimates of the noise for long signals \cite{Chatziioannou_2019}. A better approach is to estimate the noise spectrum ``on-source'' using parametrized models \cite{PhysRevD.91.084034, Cornish_2015} and to marginalise over the noise estimation uncertainty \cite{Biscoveanu_2020}.

\subsection{Non-stationary noise in LIGO and Virgo data}
\begin{figure}
  \centering
    \includegraphics[width=0.5\textwidth]{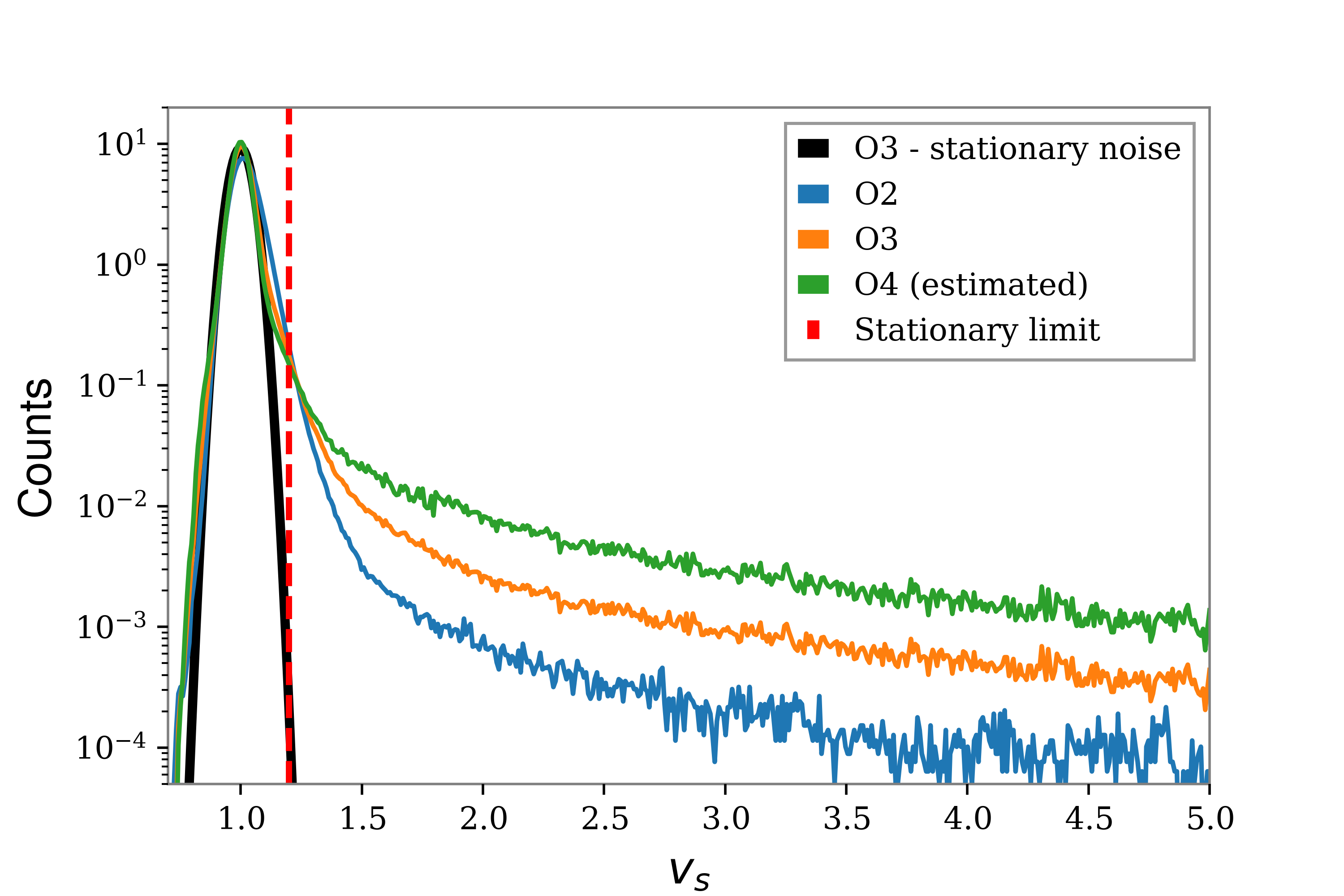}
\caption{\label{Fig:psdvar} PSD variation ($v_{s}$) distribution in LIGO Livingston based on data for O2 (blue), O3 (orange) and, estimated, O4 (green). The black curve shows the expected distribution in Gaussian stationary noise based on O3 data. The vertical red dashed line shows the limit over which we consider data to be non-stationary.}
\end{figure}

To obtain an accurate measurement of $H_0$, the standard sirens method requires us to combine several multi-messenger detections of binary neutron stars. We saw that non-stationary noise can affect the parameter estimation for longer signals. Now we want to estimate how many signals could be detected, on average, in non-stationary data.  

We identify non-stationary noise using the approach described in Ref.~\cite{Mozzon_2020}. This method relies on modelling the relation between the noise spectrum computed over a short stretch of data (typically 8 seconds) and a longer segment of  time (512 seconds) with a frequency independent factor, $v_s$, such that $S_n(short) = v_s S_n(long)$. The time series $v_s(t)$, also called PSD variation statistic, has been proven effective to track non-stationarity in LIGO and Virgo data during the third observing run (O3)~\cite{Davis_2021}. The PSD variation at each time depends only on the amplitude of the non-stationarity and is completely independent of the shape of the noise. 

In Gaussian and stationary noise, the PSD variation statistic is well modelled by a Gaussian distribution with mean 1 and variance dependent on the bandwidth of the detector, as shown by the black curve in Figure~\ref{Fig:psdvar} based on O3 data. As shown in Figure \ref{Fig:psdvar}, non-stationarity appears as a tail of high $v_s$ values. For simplicity, here we consider $v_s>1.2$ as an indicator of non-stationarity in the data. With this approximation we analyse LIGO and Virgo data. We found that the fraction of non-stationary data in the LIGO detectors almost doubles between the second observing run (O2) and O3a. During O3a $\sim$2\% of LIGO Hanford and LIGO Livingston data and $\sim$1\% of Virgo data are non-stationary. Randomly placing 10,000 signals of 128 seconds in duration in the data, we found that 15\% of the signals would lie in non-stationary noise in at least one detector for 10 or more seconds around the merger time. Therefore, an average of more than 1 in 7 BNS detections could have been affected by non-stationary noise during O3a.

Assuming the fraction of non-stationary data to be linearly dependent with the sensitivity of the detectors we predict the levels of non-stationarity in LIGO Livingston for the next two observing runs~\cite{Abbott_2020_1}. The assumption is consistent with the rate of non-stationarity observed in LIGO data from the first three observing runs. We predict that on average 4\% and 9\% of data will be non-stationary respectively for O4 and O5. Figure \ref{Fig:psdvar} shows the measured distribution of the PSD variation of LIGO Livingston for O2, O3a and the estimation for O4. For O4, we assume an average BNS range of 180 Mpc. Similar values can be predicted for LIGO Hanford, showing that considering non-stationarity will be increasingly important in the future.

\section{Estimating the effect of non-stationary noise}\label{Section:Method}

 \begin{figure*}
   \centering
     \includegraphics[width=\textwidth]{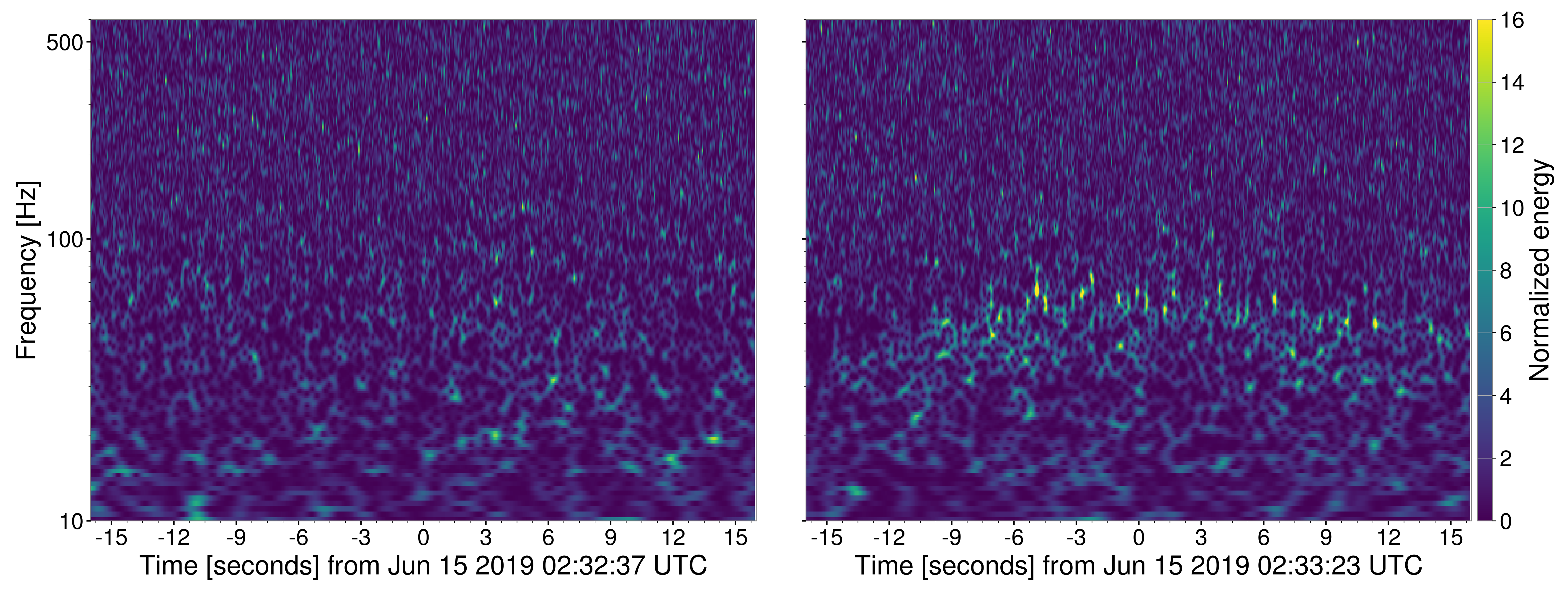}
 \caption{\label{Fig:real_times} Time-frequency spectrogram of stationary (left) and non-stationary data (right) measured by LIGO Livingston during the LIGO and Virgo third observing run.}
 \end{figure*}

In order to investigate the effect of non-stationary noise in the estimation of the luminosity distance we add a population of 467 simulated binary neutron stars to O3a LIGO and Virgo data. We target 28 separate periods of non stationary data in LIGO Livingston, with varying duration between 25 and 200 seconds. In the selected segments, we require the data in LIGO Hanford and Virgo to be stationary. In fact, coincident periods of non-stationarity are rare for ground-based detectors, representing less then 2\% of the total non-stationarity time. In future observing runs we expect more periods of coincident non-stationarity due to the larger fraction of non-stationary data; although it is unlikely this will represent the dominant scenario.

From the simulated signals we randomly select 100 signals with network signal-to-noise ratio (SNR) greater than 12 and we perform a Bayesian analysis to estimate the luminosity distance. We choose this detection threshold in order to analyse signals which are confidently detected by search pipelines even in non-stationary noise. This choice is also consistent with similar analyses \cite{Abbott:2019yzh, arxiv.2204.03614}. Finally, we compare the results of an equivalent analysis made over stationary noise close in time (but not overlapping) with the targeted non-stationarity segments. The detectors sensitivity to gravitational-wave signals varies considerably during O3a due to adjustments in the configuration of the interferometers. Considering adjacent times reduces the possibility of any variation which could affect our investigation.

We targeted moderate non-stationary noise with a PSD variation value between 1.2-3, which constitutes 80\% of all non-stationarity. Higher values generally indicate extreme non-stationarity or very short bursts of excess power \cite{TheLIGOScientific:2016zmo,2018RSPTA.37670286N,Davis_2021} that are likely to be identified and removed before performing the parameter estimation analysis. Figure \ref{Fig:real_times} shows two time-frequency spectrograms of LIGO Livingston data for one targeted time and its adjacent closest period of stationary noise. Non-stationarity appears as power excess of unknown origin distributed around 50 Hz.

\subsection{Simulations}\label{Section:Injection}
We simulate a population of non-spinning binary neutron stars with detector-frame chirp mass $\mathcal{M^{\mathrm{det}}}$ \cite{Sathyaprakash_2009} uniformly distributed between 1.7 and 1.9 $M_\odot$ and a mass ratio between $0.75-1$. We distribute mergers uniformly in Euclidean volume, extracting the signals from a prior in luminosity distance $\pi(d_L)\propto d_L^2$ \cite{PhysRevLett.116.241102, PhysRevX.9.031040} between 20 and 400~Mpc.
This approximation is appropriate to describe the observed population of BNS in the luminosity range considered \cite{10.1093/mnras/staa2850}. To reduce the computational cost we neglect tidal effects; tidal parameters do not contribute to the waveform amplitude and are not correlated with the luminosity distance. The choice of non-spinning injection is justified by the expected low number of events with high spins (e.g.~\cite{Zhu:2017znf}). Moreover, none of the BNS signals detected so far by the LIGO and Virgo Collaborations were consistent with high spins (e.g.~\cite{PhysRevX.11.021053}). We generate the signals using the waveform model~\texttt{IMRPhenomPv2} \cite{PhysRevD.91.024043, PhysRevLett.113.151101} with a low frequency cut off of 20 Hz. This model has a low computational cost and provides a good approximation for a BNS system if the tidal effects are neglected. We fix the sky position of the signals to the optimal location for our targeted detector, i.e. on the zenith for LIGO Livingston. With this choice each signal has the highest SNR in the detector which presents non-stationary noise, therefore the effect of non-stationarity on the detection is maximised. 

We injected the signals in individual stretches of data and we separate the merger times between simulations by 4 seconds. This is to avoid correlation between the estimations \cite{pizzati2021bayesian, Samajdar_2021, relton2021parameter}. We calculate the PSD using the off-source approach as described in Ref \cite{Allen_2012}, using the Welch method for 1024 seconds of data. Despite being sub-optimal compared to the ``on-source'' method, this approach is much faster for longer signals and is therefore preferable for population studies. We then use \texttt{Bilby} with the \texttt{Dynesty} sampler \cite{2020MNRAS.493.3132S} to estimate the parameters of the injections. For each signal, we analyse 128 seconds of data using~\texttt{IMRPhenomPv2} model in its reduced order quadrature approximation to reduce the computational cost \cite{Canizares_2015, Smith_2016}. We use priors consistent with the generated population of signals. Assuming the EM counterpart would allow us to uniquely identify the host galaxy, we fix the sky location to the injected value. While this is an optimistic scenario, this assumption drastically improves the accuracy in the estimation of the luminosity distance, helping to isolate and highlight the effect of non-stationarity. We assume the EM counterpart does not provide any information on the binary inclination angle.

\subsection{Bias in luminosity distance}\label{Section:Results}

\begin{figure}
  \centering
    \includegraphics[width=0.5\textwidth]{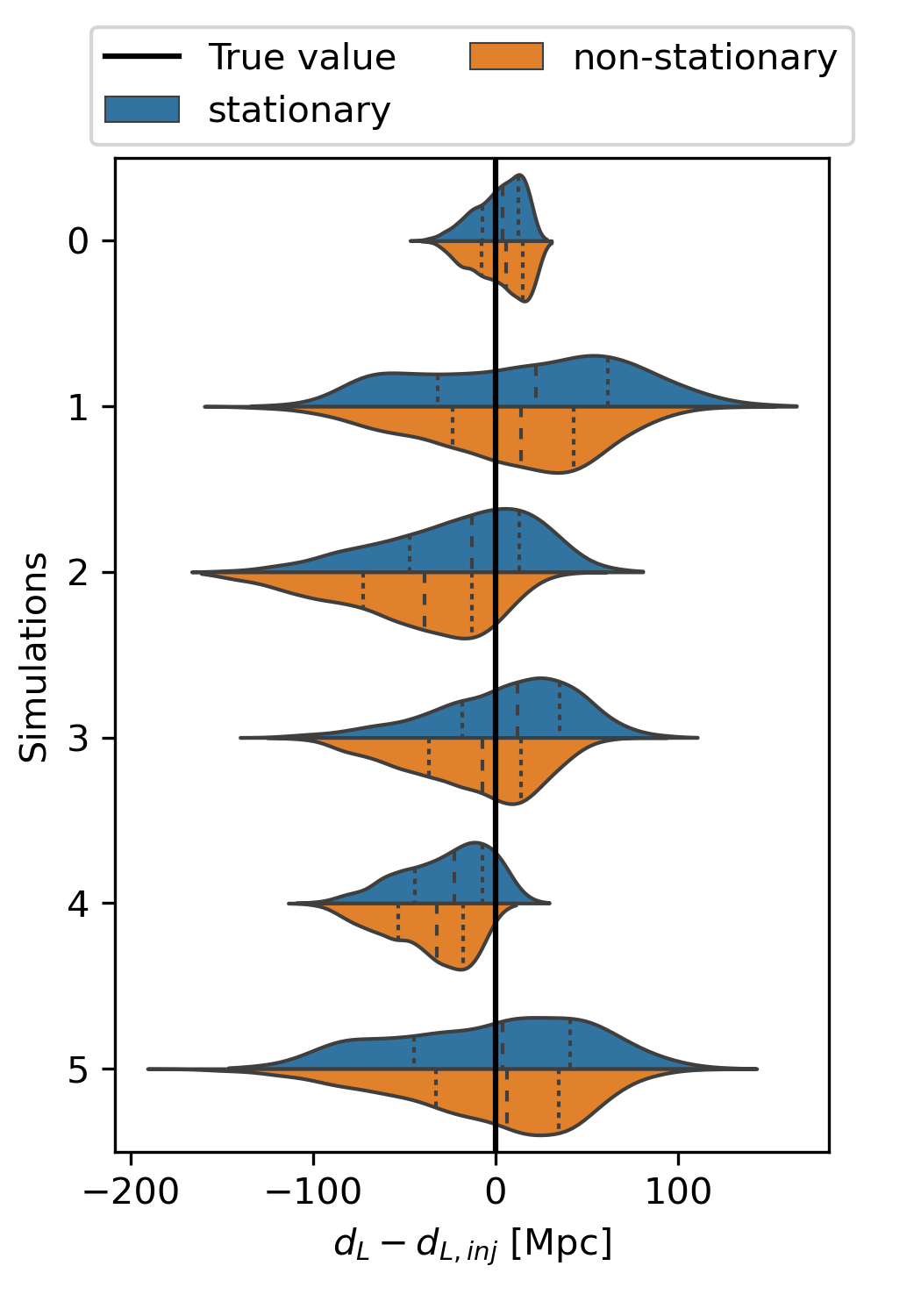}
\caption{\label{Fig:violins} Luminosity distance posteriors for 6 simulated signals added in stationary (blue) and non-stationary noise (orange). Each posterior is centred around the true value of the simulated signal. 
The dashed lines show the quartiles of the distributions.}
\end{figure}

\begin{figure}
  \centering
    \includegraphics[width=0.5\textwidth]{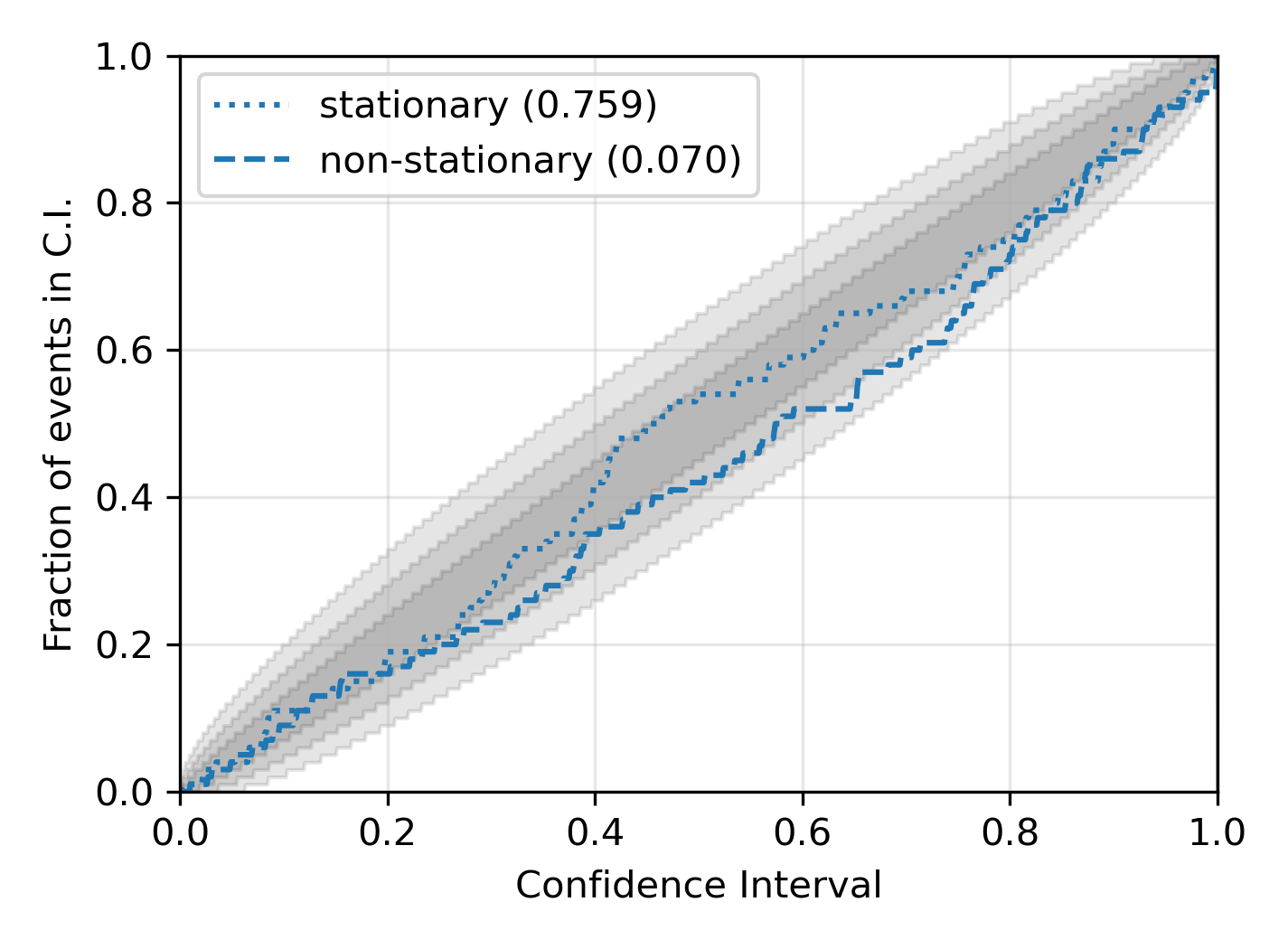}
\caption{\label{Fig:results} Results of 100 injections recovered in stationary and non-stationary noise. The grey regions cover the cumulative 1,2 and 3 $\sigma$ confidence intervals accounting for sampling errors. The blue lines represent the cumulative fraction of real luminosity distances found within this confidence interval (C.I.). Luminosity distance p-values for stationary and non-stationary noise are displayed in parentheses in the plot legend. The luminosity distance p-value for stationary noise, is 0.759, consistent with the p-value being drawn from a uniform distribution.}
\end{figure}

\begin{figure*}
  \centering
    \includegraphics[width=\textwidth]{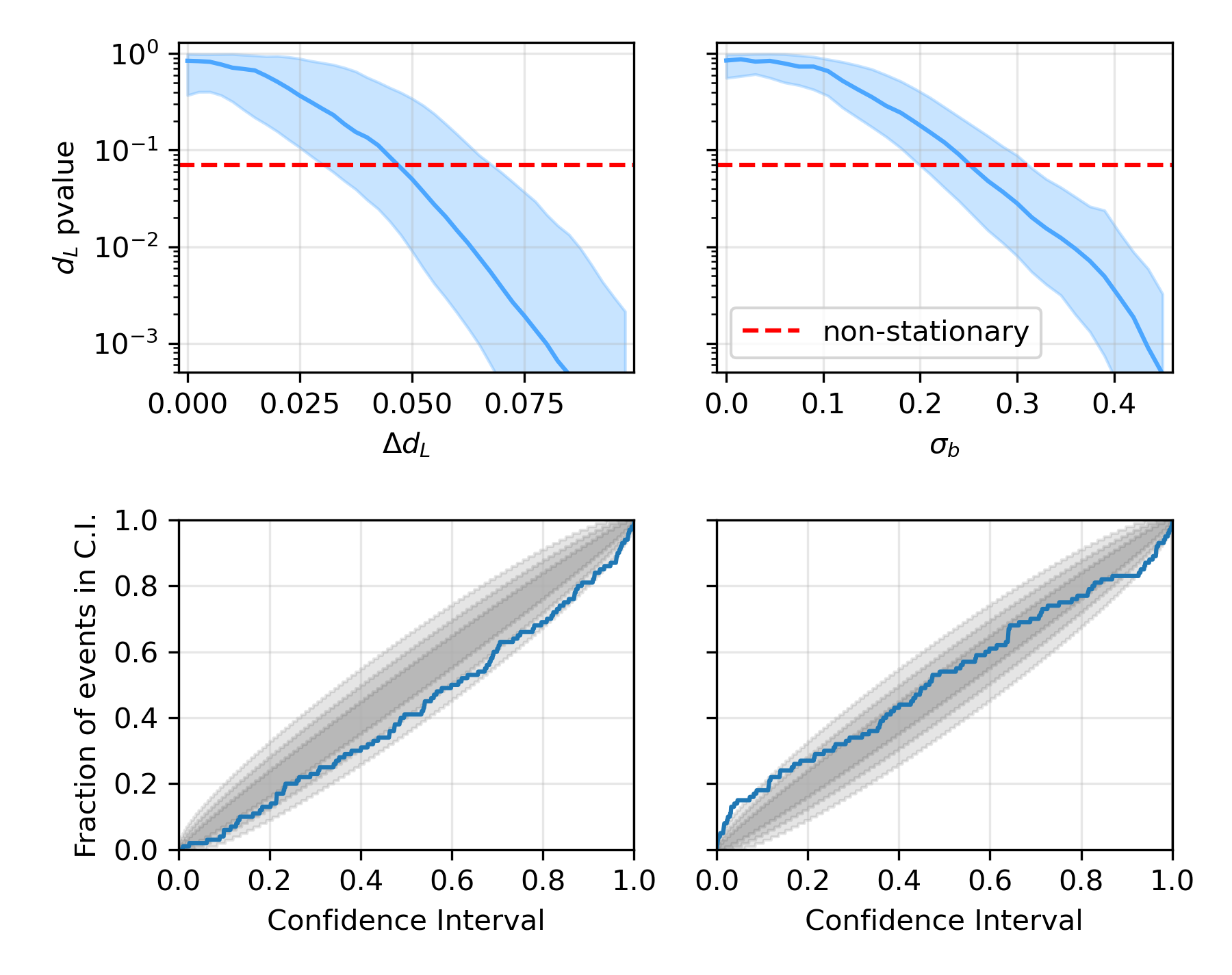}
\caption{\label{Fig:bias} Results for 100 BNS injections in simulated Gaussian noise. The top plots show the variation of the luminosity distance p-value as a function of the bias introduced in the luminosity distance posterior samples. 
For each level of bias, we calculate the Kolmogorov-Smirnov p-value for 100 posteriors randomly selected from our sample of signals. The blue line represents the median luminosity distance p-value from 50 different random samplings. The blue area delimits the 5th and the 95th percentiles of the p-value distribution.
The red dashed line shows the p-value measured for events in non-stationary noise. In the bottom plots we show how the bias distorts the cumulative fraction of injected luminosity distances found within the confidence interval in two particular cases: for $\Delta d_L = 0.047$ (\textit{Left}) and $\sigma_b = 0.25$ (\textit{Right}). These values correspond to the intersections between the blue and red lines in the upper plot, i.e. the median biases associated with the non-stationary p-values.}
\end{figure*}

We first present the luminosity distance posterior distributions for a representative sample of the simulated signals. 
In Figure \ref{Fig:violins} we compare distributions obtained for identical signals detected in stationary and non-stationary noise. The effects of non-stationarity appear to vary for different signals. While signals labelled as 1 and 5 appear to be over-constrained in non-stationary noise, the main effect on simulations 2, 3 and 4 is a shift towards smaller luminosity distance values. 
Considering signals detected in both stationary and non-stationary, we found that the median distance is reduced on average by 1.4\%. Similarly, the 25th and the 75th percentiles of the estimated luminosity distance distribution are shifted on average by 1.1\% and 1.5\%, suggesting non-stationarity might cause a rigid shift of the luminosity distance posteriors. Note that only 80\% of signals are detected in both stationary and non-stationary noise. The remaining detections vary between the two sets.
Considering all the signals, we found the median of the luminosity distance posteriors to be lower than the correspondent true value for 58\% of the signals in non-stationary noise, in contrast with the 47\% for signals in stationary noise. These differences indicate the possible presence of systematics in the estimations in non-stationary noise. However, directly comparing the posterior distributions obtained with and without non-stationary noise for each event is not sufficient to identify systematic biases. In fact, different realisations of stationary Gaussian noise might also cause the inferred parameters to vary.


To verify if the observed distortions could be explained as random variations of the noise, we compare the luminosity distance posteriors computing the normalised cumulative fraction of true luminosity distances which lie within a measured confidence interval \cite{doi:10.1198/106186006X136976}. This approach is commonly used to identify biases in the inference on gravitational-wave sources \cite{10.1093/mnras/staa2850, gair2015quantifying, Berry_2015, Biwer_2019, PhysRevD.89.084060,  PhysRevD.92.023002}.
 
Figure \ref{Fig:results} shows the results for 100 injections in stationary and non-stationary noise. This is generally referred to as a P-P plot.
If the inference is unbiased the fraction of events in a particular confidence interval is drawn from a uniform distribution. Hence, the expected cumulative distribution would lie on the diagonal of the plot with some scatter due to Poisson error. The shaded regions delimit the expected 1, 2 and 3-sigma error given the number of events. For signals injected in non-stationary noise the cumulative distribution is systematically below the diagonal, exceeding the 2-sigma error for confidence intervals between 0.6 and 0.8. 

We test the consistency between the measured curves and the diagonal line using the Kolmogorov-Smirnov (KS) statistic. For unbiased parameter estimations the two-tailed p-value is uniformly distributed between 0 and 1. Therefore, a p-value $<0.05$ will occur once in 20 times. 
For the curves in Figure \ref{Fig:results} the resulting p-values are $0.759$ and $0.070$ for stationary and non-stationary noise respectively. 
A smaller p-value indicates that the measured curve is unlikely to be randomly extracted from the assumed distribution if the sampler is unbiased. In particular, there is just a 7\% chance to obtain a more extreme curve than the one measured from events in non-stationary noise. As shown in Table \ref{snr table}, we obtained higher p-values when increasing the cut in SNR, showing that the distortion is reduced for louder signals. However, even for louder signals the PP-plot presents similarities with Figure \ref{Fig:results}.

\begin{table}
 \begin{center}
  \begin{tabular}{||c || c | c | c||} 
    \hline
    SNR cut & 12 & 13 & 14 \\ [0.5ex] 
    \hline\hline
    stationary & 0.759 & 0.535 & 0.468 \\
    \hline
    non-stationary & 0.070 & 0.161 & 0.368 \\
    \hline
  \end{tabular}
  \caption{\label{snr table} Luminosity distance p-values for stationary and non-stationary noise as a function of the SNR cut imposed.}
 \end{center}
\end{table}

The inconsistencies in the observed p-values can be interpreted as a systematic bias in the estimated luminosity distance in non-stationary noise. Power excess in the data can increase the matched-filter SNR of the detection, decreasing the estimated luminosity distance. However, a lower p-value can also arise from over-constraining the posterior. If the posterior is over-constrained we would see a larger fraction of events in a lower confidence intervals and smaller fraction for higher intervals. This distortion with respect to the predicted curve would lower the p-value. 

To investigate these two scenarios and quantify the bias, we repeat our analysis adding 800 signals in simulated Gaussian stationary noise. For consistency with the analysis in real data we require a network SNR$>$12, which yields a final sample of 153 signals. For each signal we then artificially bias the estimated luminosity distance posterior $d_L$ by shifting the distribution by a constant value $\Delta d_L$, such that: 
\begin{equation}
    d_{L, \mathrm{biased}} = d_{L} - \Delta d_L\times d_{L,inj}
\end{equation}
where $d_{L,inj}$ is the injected luminosity distance. We calculate the K-S p-value randomly selecting 100 signals from our sample. To consider variations in the estimated p-value due to selection effects, we repeat the calculation for 50 different random samples of signals. 
Finally, we investigate the relation between the p-value and the bias repeating this procedure for increasing values of $\Delta d_L$, re-estimating the p-value for each iteration. 

We perform a similar test to understand the effect of over-constraining the posterior. In this case we uniformly shrink the distribution around the median luminosity distance. This modification has the effect to reduce the samples standard deviation of the luminosity distance distribution $\sigma(d_L)$ by a factor $\sigma_b$ for each signal. For example, a $\sigma_b = 0.1$ refers to a 10\% reduction of the standard deviation of the luminosity distance distribution for each signal. 

The top plots of Figure \ref{Fig:bias} show the evolution of the luminosity distance p-value as a function of the bias introduced in the posteriors for the two cases. The blue line represents the median p-values. for each level of bias. The blue area enclose all values between the 5th and the 95th percentiles.
The relation between the p-value and the bias is monotonic, with lower p-values indicating greater biases. Indeed, greater distortions of the posterior distribution makes the assumption that the sampler is unbiased more unlikely. 
From these plots we can infer the bias associated with the p-value observed for events in non-stationary noise, which is indicated in Figure \ref{Fig:bias} with a red dashed line. The measured p-value is consistent with a $4.7^{+2.1}_{-1.7}$\% systematic under-estimation of the measured luminosity distance or a $25^{+6}_{-5}$\% reduction in the dispersion of the posteriors distribution. 
In the bottom plots of Figure \ref{Fig:bias} we show how these two biases distort the P-P plots. Reducing the samples variance (right plot) induces an ``S-shape" in the cumulative fraction of injections found in each confidence interval, increasing it for lower confidence intervals and decreasing it for higher levels. Instead, reducing the mean of the posteriors (left plot) systematically decreases the cumulative fraction of real luminosity distances in each confidence interval. Qualitatively comparing these plots with Figure \ref{Fig:results}, we can conclude that the measured dominant effect of non-stationary noise is a systematic under-estimation of the luminosity distance.


\subsection{Considerations on the estimation of $H_0$}
In principle, a 4.7\% systematic under-estimation of the luminosity distance of the source combined with the other expected systematic uncertainties could dramatically affect the accuracy of the estimation of $H_0$ using standard sirens. Despite this inaccuracy, standard sirens would still help to break the $H_0$ tension. Let us consider the worst case, in which the systematics due to non-stationarity and calibration simply add up. Considering a 1\% calibration error, this would correspond to a 5.7\% systematic under-estimation of the luminosity distance, i.e. we would infer 5.7\% higher values of $H_0$. Assuming the early universe estimation of $H_0$ ($67.4\pm 0.5$ km~s$^{-1}$~Mpc$^{-1}$ \cite{refId0}) to be correct, we would obtain $H_0=71.2\pm 0.7$ km~s$^{-1}$~Mpc$^{-1}$, in which we assumed our estimation to be Gaussian distributed with a 1\% error. In this case, the effect of non-stationary noise would make the standard sirens estimation to fall between the early universe measurement and the local distance ladder estimation of $H_0$ ($74.03\pm 1.42$ km~s$^{-1}$~Mpc$^{-1}$ \cite{Riess_2019}). Therefore, none of the two hypothesis would be confidently excluded. In the worst case presented in Figure \ref{Fig:bias}, i.e. a 6.8\% under-estimation of the luminosity distance, the standard sirens method could also favour the wrong hypothesis.  

However, our results represent an upper limit in the shift of the luminosity distance due to non-stationary noise, with the measured p-value likely to be a result of various effects. 
Moreover, to reach the precision of a few percent required to solve the current tension on the estimation of $H_0$ it may be necessary to combine at least $\sim$50 standard sirens~\cite{Feeney_2019}. Of them, just a fraction will be affected by non-stationary noise, hence the error on the estimation will be reduced. 
On the other hand, the number of standard sirens required to attain the necessary precision may be reduced by additional EM constraints on e.g., source orientation, as with GW170817~\cite{Hotokezaka:2018dfi,Mukherjee:2019qmm}, and this could further compound any present bias.
Therefore, assessing the level of non-stationarity for individual BNS detections will be important in confidently presenting estimates of $H_0$ free from significant bias.

Other methods to estimate $H_0$ which rely on shorter signals like binary black holes will also be important to improve the accuracy on $H_0$~\cite{Fishbach_2019, Soares_Santos_2019, Gray_2020, Finke_2021}. These approaches require shorter periods of data, therefore the effect of non-stationary noise will be less important.

\section{Summary and Conclusions}\label{Section:Conclusions}
In this paper we have investigated whether the presence of non-stationarity in LIGO and Virgo data introduces a new source of systematic error in the estimation of the Hubble constant through the standard sirens approach. The problem is particularly important for longer duration signals such as BNS, for which longer periods of data are required, making the parameter estimation more vulnerable to fluctuations in the detector noise. Indeed, we found that during O3a non-stationarity accounts for 2\% of the overall LIGO data. By placing simulated BNS signals of 128 seconds in length throughout O3a data, we find that 1 in 7 BNS signals merger times could have fallen in non-stationary noise. More importantly, this fraction is predicted to increase, with non-stationarity that will reach an estimated 4\% and 9\% of the overall data for O4 and O5 respectively.

We explore the issue of non-stationarity and how it affects the estimation of the luminosity distance of the source, by adding simulated BNS signals in stationary and non-stationary data from O3a. We compare the luminosity distance posteriors obtained in the two cases calculating the cumulative fraction of true luminosity distances which lie within a measured confidence interval. We employ the Kolmogorov-Smirnov test to estimate the consistency of the results with the theoretical predictions. We found a lower p-value (0.070) for events in non-stationary noise, showing that the null hypothesis of an unbiased estimation is unlikely. In order to understand the magnitude of the misestimation, we artificially bias the posteriors of BNS signals estimated in simulated Gaussian and stationary noise. We found that the measured p-value for non-stationary noise is consistent with a systematic under-estimation of the measured luminosity distance by up to 6.8\%. 

The estimated bias in the luminosity distance is an upper limit and does not automatically translate to an expected systematic error in the estimation of $H_0$. First, just a fraction of the BNS-like gravitational-wave detections will be measured in non-stationary noise. It is estimated that $\mathcal{O}(100)$ joint gravitational-wave and EM detections are needed in order to infer $H_0$ to an accuracy of 1$\%$. Therefore the combination of these $\mathcal{O}(100)$ signals, of which $\sim 15\%$ may be affected by non-stationarity, is unlikely to have a large effect on the accuracy of $H_0$. Moreover, binary black holes detections are expected to give an important contribution to improve the accuracy of $H_0$ \cite{LVK:2021bbr}. The duration of these signals is on the order of seconds, making the effect of non-stationary noise less important. 
Therefore, despite non-stationarity in LIGO and Virgo data will affect the standard sirens estimation of $H_0$, we do not expect it to be a limiting factor in resolving the tension on $H_0$ using data from second generation (2G) detectors.
However, until gravitational-wave inference methods fully account for non-stationary noise, assessing the level of non-stationarity in the data, in particular for louder signals, will be crucial to exclude biases in the $H_0$ estimation. 


The next generation (3G) of gravitational-wave detectors, such as the Einstein Telescope~\cite{Punturo:2010zz,Maggiore:2019uih} and Cosmic Explorer~\cite{LIGOScientific:2016wof,Reitze:2019iox}, with their increased sensitivity at lower frequencies, will detect much longer duration gravitational-wave signals than current 2G detectors. Non-stationarity will still be an issue in these detectors, although we can not estimate, just yet, whether it will be at similar levels or worse than what we see in the 2G detectors. Either way, non-stationarity will have to be considered in the interpretation of long duration signals in 3G detectors to ensure this form of noise does not impact key scientific conclusions. 

\acknowledgments
We are grateful to the referees for the very valuable comments and suggestions. 
We are thankful to the LIGO/Virgo Cosmology group for insightful comments on this work, as well as Sylvia Biscoveanu and Ian Harry for useful discussions on this paper.

SM was supported by a STFC studentship. GA and LKN thank the UKRI Future Leaders Fellowship for support through the
grant MR/T01881X/1. ARW thanks the STFC for support through the grant ST/S000550/1.

This research has made use of data, software and/or web tools obtained from the Gravitational Wave Open Science Center (https://www.gw-openscience.org), a service of LIGO Laboratory, the LIGO Scientific Collaboration and the Virgo Collaboration. LIGO is funded by the U.S. National Science Foundation. Virgo is funded by the French Centre National de Recherche Scientifique (CNRS), the Italian Istituto Nazionale della Fisica Nucleare (INFN) and the Dutch Nikhef, with contributions by Polish and Hungarian institutes. The  authors  are  grateful for  computational resources  provided  by  the  LIGO  Laboratory  and  supported  by  National  Science  Foundation  Grants  PHY-0757058 and PHY-0823459. 

This work carries LIGO Document number P2100316.

\bibliography{references}

\begin{thebibliography}{89}%
\makeatletter
\providecommand \@ifxundefined [1]{%
 \@ifx{#1\undefined}
}%
\providecommand \@ifnum [1]{%
 \ifnum #1\expandafter \@firstoftwo
 \else \expandafter \@secondoftwo
 \fi
}%
\providecommand \@ifx [1]{%
 \ifx #1\expandafter \@firstoftwo
 \else \expandafter \@secondoftwo
 \fi
}%
\providecommand \natexlab [1]{#1}%
\providecommand \enquote  [1]{``#1''}%
\providecommand \bibnamefont  [1]{#1}%
\providecommand \bibfnamefont [1]{#1}%
\providecommand \citenamefont [1]{#1}%
\providecommand \href@noop [0]{\@secondoftwo}%
\providecommand \href [0]{\begingroup \@sanitize@url \@href}%
\providecommand \@href[1]{\@@startlink{#1}\@@href}%
\providecommand \@@href[1]{\endgroup#1\@@endlink}%
\providecommand \@sanitize@url [0]{\catcode `\\12\catcode `\$12\catcode
  `\&12\catcode `\#12\catcode `\^12\catcode `\_12\catcode `\%12\relax}%
\providecommand \@@startlink[1]{}%
\providecommand \@@endlink[0]{}%
\providecommand \url  [0]{\begingroup\@sanitize@url \@url }%
\providecommand \@url [1]{\endgroup\@href {#1}{\urlprefix }}%
\providecommand \urlprefix  [0]{URL }%
\providecommand \Eprint [0]{\href }%
\providecommand \doibase [0]{http://dx.doi.org/}%
\providecommand \selectlanguage [0]{\@gobble}%
\providecommand \bibinfo  [0]{\@secondoftwo}%
\providecommand \bibfield  [0]{\@secondoftwo}%
\providecommand \translation [1]{[#1]}%
\providecommand \BibitemOpen [0]{}%
\providecommand \bibitemStop [0]{}%
\providecommand \bibitemNoStop [0]{.\EOS\space}%
\providecommand \EOS [0]{\spacefactor3000\relax}%
\providecommand \BibitemShut  [1]{\csname bibitem#1\endcsname}%
\let\auto@bib@innerbib\@empty
\bibitem [{\citenamefont {{Schutz}}(1986)}]{Schutz:1986ss}%
  \BibitemOpen
  \bibfield  {author} {\bibinfo {author} {\bibfnamefont {B.~F.}\ \bibnamefont
  {{Schutz}}},\ }\href {\doibase 10.1038/323310a0} {\bibfield  {journal}
  {\bibinfo  {journal} {\nat}\ }\textbf {\bibinfo {volume} {323}},\ \bibinfo
  {pages} {310} (\bibinfo {year} {1986})}\BibitemShut {NoStop}%
\bibitem [{\citenamefont {Holz}\ and\ \citenamefont
  {Hughes}(2005)}]{Holz_2005}%
  \BibitemOpen
  \bibfield  {author} {\bibinfo {author} {\bibfnamefont {D.~E.}\ \bibnamefont
  {Holz}}\ and\ \bibinfo {author} {\bibfnamefont {S.~A.}\ \bibnamefont
  {Hughes}},\ }\href {\doibase 10.1086/431341} {\bibfield  {journal} {\bibinfo
  {journal} {The Astrophysical Journal}\ }\textbf {\bibinfo {volume} {629}},\
  \bibinfo {pages} {15–22} (\bibinfo {year} {2005})}\BibitemShut {NoStop}%
\bibitem [{\citenamefont {Dalal}\ \emph {et~al.}(2006)\citenamefont {Dalal},
  \citenamefont {Holz}, \citenamefont {Hughes},\ and\ \citenamefont
  {Jain}}]{Dalal_2006}%
  \BibitemOpen
  \bibfield  {author} {\bibinfo {author} {\bibfnamefont {N.}~\bibnamefont
  {Dalal}}, \bibinfo {author} {\bibfnamefont {D.~E.}\ \bibnamefont {Holz}},
  \bibinfo {author} {\bibfnamefont {S.~A.}\ \bibnamefont {Hughes}}, \ and\
  \bibinfo {author} {\bibfnamefont {B.}~\bibnamefont {Jain}},\ }\href {\doibase
  10.1103/PhysRevD.74.063006} {\bibfield  {journal} {\bibinfo  {journal} {Phys.
  Rev. D}\ }\textbf {\bibinfo {volume} {74}},\ \bibinfo {pages} {063006}
  (\bibinfo {year} {2006})}\BibitemShut {NoStop}%
\bibitem [{\citenamefont {Nissanke}\ \emph {et~al.}(2010)\citenamefont
  {Nissanke}, \citenamefont {Holz}, \citenamefont {Hughes}, \citenamefont
  {Dalal},\ and\ \citenamefont {Sievers}}]{Nissanke_2010}%
  \BibitemOpen
  \bibfield  {author} {\bibinfo {author} {\bibfnamefont {S.}~\bibnamefont
  {Nissanke}}, \bibinfo {author} {\bibfnamefont {D.~E.}\ \bibnamefont {Holz}},
  \bibinfo {author} {\bibfnamefont {S.~A.}\ \bibnamefont {Hughes}}, \bibinfo
  {author} {\bibfnamefont {N.}~\bibnamefont {Dalal}}, \ and\ \bibinfo {author}
  {\bibfnamefont {J.~L.}\ \bibnamefont {Sievers}},\ }\href {\doibase
  10.1088/0004-637x/725/1/496} {\bibfield  {journal} {\bibinfo  {journal} {The
  Astrophysical Journal}\ }\textbf {\bibinfo {volume} {725}},\ \bibinfo {pages}
  {496} (\bibinfo {year} {2010})}\BibitemShut {NoStop}%
\bibitem [{\citenamefont {{Aghanim, N.}}\ \emph {et~al.}(2020)\citenamefont
  {{Aghanim, N.}} \emph {et~al.}}]{refId0}%
  \BibitemOpen
  \bibfield  {author} {\bibinfo {author} {\bibnamefont {{Aghanim, N.}}} \emph
  {et~al.} (\bibinfo {collaboration} {Planck Collaboration}),\ }\href {\doibase
  10.1051/0004-6361/201833910} {\bibfield  {journal} {\bibinfo  {journal}
  {A\&A}\ }\textbf {\bibinfo {volume} {641}},\ \bibinfo {pages} {A6} (\bibinfo
  {year} {2020})}\BibitemShut {NoStop}%
\bibitem [{\citenamefont {Riess}\ \emph {et~al.}(2019)\citenamefont {Riess},
  \citenamefont {Casertano}, \citenamefont {Yuan}, \citenamefont {Macri},\ and\
  \citenamefont {Scolnic}}]{Riess_2019}%
  \BibitemOpen
  \bibfield  {author} {\bibinfo {author} {\bibfnamefont {A.~G.}\ \bibnamefont
  {Riess}}, \bibinfo {author} {\bibfnamefont {S.}~\bibnamefont {Casertano}},
  \bibinfo {author} {\bibfnamefont {W.}~\bibnamefont {Yuan}}, \bibinfo {author}
  {\bibfnamefont {L.~M.}\ \bibnamefont {Macri}}, \ and\ \bibinfo {author}
  {\bibfnamefont {D.}~\bibnamefont {Scolnic}},\ }\href {\doibase
  10.3847/1538-4357/ab1422} {\bibfield  {journal} {\bibinfo  {journal} {The
  Astrophysical Journal}\ }\textbf {\bibinfo {volume} {876}},\ \bibinfo {pages}
  {85} (\bibinfo {year} {2019})}\BibitemShut {NoStop}%
\bibitem [{\citenamefont {Di~Valentino}\ \emph {et~al.}(2021)\citenamefont
  {Di~Valentino}, \citenamefont {Mena}, \citenamefont {Pan}, \citenamefont
  {Visinelli}, \citenamefont {Yang}, \citenamefont {Melchiorri}, \citenamefont
  {Mota}, \citenamefont {Riess},\ and\ \citenamefont
  {Silk}}]{DiValentino:2021izs}%
  \BibitemOpen
  \bibfield  {author} {\bibinfo {author} {\bibfnamefont {E.}~\bibnamefont
  {Di~Valentino}}, \bibinfo {author} {\bibfnamefont {O.}~\bibnamefont {Mena}},
  \bibinfo {author} {\bibfnamefont {S.}~\bibnamefont {Pan}}, \bibinfo {author}
  {\bibfnamefont {L.}~\bibnamefont {Visinelli}}, \bibinfo {author}
  {\bibfnamefont {W.}~\bibnamefont {Yang}}, \bibinfo {author} {\bibfnamefont
  {A.}~\bibnamefont {Melchiorri}}, \bibinfo {author} {\bibfnamefont {D.~F.}\
  \bibnamefont {Mota}}, \bibinfo {author} {\bibfnamefont {A.~G.}\ \bibnamefont
  {Riess}}, \ and\ \bibinfo {author} {\bibfnamefont {J.}~\bibnamefont {Silk}},\
  }\href {\doibase 10.1088/1361-6382/ac086d} {\bibfield  {journal} {\bibinfo
  {journal} {Class. Quant. Grav.}\ }\textbf {\bibinfo {volume} {38}},\ \bibinfo
  {pages} {153001} (\bibinfo {year} {2021})},\ \Eprint
  {http://arxiv.org/abs/2103.01183} {arXiv:2103.01183 [astro-ph.CO]}
  \BibitemShut {NoStop}%
\bibitem [{\citenamefont {Freedman}(2021)}]{Freedman:2021ahq}%
  \BibitemOpen
  \bibfield  {author} {\bibinfo {author} {\bibfnamefont {W.~L.}\ \bibnamefont
  {Freedman}},\ }\href {\doibase 10.3847/1538-4357/ac0e95} {\bibfield
  {journal} {\bibinfo  {journal} {The Astrophysical Journal}\ }\textbf
  {\bibinfo {volume} {919}},\ \bibinfo {pages} {16} (\bibinfo {year}
  {2021})}\BibitemShut {NoStop}%
\bibitem [{\citenamefont {Abbott}\ \emph
  {et~al.}(2017{\natexlab{a}})\citenamefont {Abbott} \emph
  {et~al.}}]{LIGOScientific:2017adf}%
  \BibitemOpen
  \bibfield  {author} {\bibinfo {author} {\bibfnamefont {B.~P.}\ \bibnamefont
  {Abbott}} \emph {et~al.} (\bibinfo {collaboration} {LIGO Scientific, Virgo,
  1M2H, Dark Energy Camera GW-E, DES, DLT40, Las Cumbres Observatory, VINROUGE,
  MASTER}),\ }\href {\doibase 10.1038/nature24471} {\bibfield  {journal}
  {\bibinfo  {journal} {Nature}\ }\textbf {\bibinfo {volume} {551}},\ \bibinfo
  {pages} {85} (\bibinfo {year} {2017}{\natexlab{a}})},\ \Eprint
  {http://arxiv.org/abs/1710.05835} {arXiv:1710.05835 [astro-ph.CO]}
  \BibitemShut {NoStop}%
\bibitem [{\citenamefont {{Freedman}}(2017)}]{freedman2017cosmology}%
  \BibitemOpen
  \bibfield  {author} {\bibinfo {author} {\bibfnamefont {W.~L.}\ \bibnamefont
  {{Freedman}}},\ }\href {\doibase 10.1038/s41550-017-0121} {\bibfield
  {journal} {\bibinfo  {journal} {Nature Astronomy}\ }\textbf {\bibinfo
  {volume} {1}},\ \bibinfo {eid} {0121} (\bibinfo {year} {2017})}\BibitemShut
  {NoStop}%
\bibitem [{\citenamefont {Nissanke}\ \emph {et~al.}(2013)\citenamefont
  {Nissanke}, \citenamefont {Holz}, \citenamefont {Dalal}, \citenamefont
  {Hughes}, \citenamefont {Sievers},\ and\ \citenamefont
  {Hirata}}]{nissanke2013determining}%
  \BibitemOpen
  \bibfield  {author} {\bibinfo {author} {\bibfnamefont {S.}~\bibnamefont
  {Nissanke}}, \bibinfo {author} {\bibfnamefont {D.~E.}\ \bibnamefont {Holz}},
  \bibinfo {author} {\bibfnamefont {N.}~\bibnamefont {Dalal}}, \bibinfo
  {author} {\bibfnamefont {S.~A.}\ \bibnamefont {Hughes}}, \bibinfo {author}
  {\bibfnamefont {J.~L.}\ \bibnamefont {Sievers}}, \ and\ \bibinfo {author}
  {\bibfnamefont {C.~M.}\ \bibnamefont {Hirata}},\ }\href@noop {} {\bibfield
  {journal} {\bibinfo  {journal} {arXiv e-prints}\ ,\ \bibinfo {eid}
  {arXiv:1307.2638}} (\bibinfo {year} {2013})},\ \Eprint
  {http://arxiv.org/abs/1307.2638} {arXiv:1307.2638 [astro-ph.CO]} \BibitemShut
  {NoStop}%
\bibitem [{\citenamefont {Chen}\ \emph {et~al.}(2018)\citenamefont {Chen},
  \citenamefont {Fishbach},\ and\ \citenamefont {Holz}}]{Chen_2018}%
  \BibitemOpen
  \bibfield  {author} {\bibinfo {author} {\bibfnamefont {H.-Y.}\ \bibnamefont
  {Chen}}, \bibinfo {author} {\bibfnamefont {M.}~\bibnamefont {Fishbach}}, \
  and\ \bibinfo {author} {\bibfnamefont {D.~E.}\ \bibnamefont {Holz}},\ }\href
  {\doibase 10.1038/s41586-018-0606-0} {\bibfield  {journal} {\bibinfo
  {journal} {Nature}\ }\textbf {\bibinfo {volume} {562}},\ \bibinfo {pages}
  {545–547} (\bibinfo {year} {2018})}\BibitemShut {NoStop}%
\bibitem [{\citenamefont {Feeney}\ \emph {et~al.}(2019)\citenamefont {Feeney},
  \citenamefont {Peiris}, \citenamefont {Williamson}, \citenamefont {Nissanke},
  \citenamefont {Mortlock}, \citenamefont {Alsing},\ and\ \citenamefont
  {Scolnic}}]{Feeney_2019}%
  \BibitemOpen
  \bibfield  {author} {\bibinfo {author} {\bibfnamefont {S.~M.}\ \bibnamefont
  {Feeney}}, \bibinfo {author} {\bibfnamefont {H.~V.}\ \bibnamefont {Peiris}},
  \bibinfo {author} {\bibfnamefont {A.~R.}\ \bibnamefont {Williamson}},
  \bibinfo {author} {\bibfnamefont {S.~M.}\ \bibnamefont {Nissanke}}, \bibinfo
  {author} {\bibfnamefont {D.~J.}\ \bibnamefont {Mortlock}}, \bibinfo {author}
  {\bibfnamefont {J.}~\bibnamefont {Alsing}}, \ and\ \bibinfo {author}
  {\bibfnamefont {D.}~\bibnamefont {Scolnic}},\ }\href {\doibase
  10.1103/PhysRevLett.122.061105} {\bibfield  {journal} {\bibinfo  {journal}
  {Phys. Rev. Lett.}\ }\textbf {\bibinfo {volume} {122}},\ \bibinfo {pages}
  {061105} (\bibinfo {year} {2019})}\BibitemShut {NoStop}%
\bibitem [{\citenamefont {Mortlock}\ \emph {et~al.}(2019)\citenamefont
  {Mortlock}, \citenamefont {Feeney}, \citenamefont {Peiris}, \citenamefont
  {Williamson},\ and\ \citenamefont {Nissanke}}]{2019PhRvD.100j3523M}%
  \BibitemOpen
  \bibfield  {author} {\bibinfo {author} {\bibfnamefont {D.~J.}\ \bibnamefont
  {Mortlock}}, \bibinfo {author} {\bibfnamefont {S.~M.}\ \bibnamefont
  {Feeney}}, \bibinfo {author} {\bibfnamefont {H.~V.}\ \bibnamefont {Peiris}},
  \bibinfo {author} {\bibfnamefont {A.~R.}\ \bibnamefont {Williamson}}, \ and\
  \bibinfo {author} {\bibfnamefont {S.~M.}\ \bibnamefont {Nissanke}},\ }\href
  {\doibase 10.1103/PhysRevD.100.103523} {\bibfield  {journal} {\bibinfo
  {journal} {Phys. Rev. D}\ }\textbf {\bibinfo {volume} {100}},\ \bibinfo
  {pages} {103523} (\bibinfo {year} {2019})}\BibitemShut {NoStop}%
\bibitem [{\citenamefont {Howlett}\ and\ \citenamefont
  {Davis}(2020)}]{10.1093/mnras/staa049}%
  \BibitemOpen
  \bibfield  {author} {\bibinfo {author} {\bibfnamefont {C.}~\bibnamefont
  {Howlett}}\ and\ \bibinfo {author} {\bibfnamefont {T.~M.}\ \bibnamefont
  {Davis}},\ }\href@noop {} {\bibfield  {journal} {\bibinfo  {journal} {Monthly
  Notices of the Royal Astronomical Society}\ }\textbf {\bibinfo {volume}
  {492}},\ \bibinfo {pages} {3803} (\bibinfo {year} {2020})}\BibitemShut
  {NoStop}%
\bibitem [{\citenamefont {Nicolaou}\ \emph {et~al.}(2020)\citenamefont
  {Nicolaou}, \citenamefont {Lahav}, \citenamefont {Lemos}, \citenamefont
  {Hartley},\ and\ \citenamefont {Braden}}]{10.1093/mnras/staa1120}%
  \BibitemOpen
  \bibfield  {author} {\bibinfo {author} {\bibfnamefont {C.}~\bibnamefont
  {Nicolaou}}, \bibinfo {author} {\bibfnamefont {O.}~\bibnamefont {Lahav}},
  \bibinfo {author} {\bibfnamefont {P.}~\bibnamefont {Lemos}}, \bibinfo
  {author} {\bibfnamefont {W.}~\bibnamefont {Hartley}}, \ and\ \bibinfo
  {author} {\bibfnamefont {J.}~\bibnamefont {Braden}},\ }\href@noop {}
  {\bibfield  {journal} {\bibinfo  {journal} {Monthly Notices of the Royal
  Astronomical Society}\ }\textbf {\bibinfo {volume} {495}},\ \bibinfo {pages}
  {90} (\bibinfo {year} {2020})}\BibitemShut {NoStop}%
\bibitem [{\citenamefont {{Mukherjee, Suvodip}}\ \emph
  {et~al.}(2021{\natexlab{a}})\citenamefont {{Mukherjee, Suvodip}},
  \citenamefont {{Lavaux, Guilhem}}, \citenamefont {{Bouchet, Fran\c{c}ois
  R.}}, \citenamefont {{Jasche, Jens}}, \citenamefont {{Wandelt, Benjamin D.}},
  \citenamefont {{Nissanke, Samaya}}, \citenamefont {{Leclercq, Florent}},\
  and\ \citenamefont {{Hotokezaka, Kenta}}}]{Mukherjee_2021}%
  \BibitemOpen
  \bibfield  {author} {\bibinfo {author} {\bibnamefont {{Mukherjee, Suvodip}}},
  \bibinfo {author} {\bibnamefont {{Lavaux, Guilhem}}}, \bibinfo {author}
  {\bibnamefont {{Bouchet, Fran\c{c}ois R.}}}, \bibinfo {author} {\bibnamefont
  {{Jasche, Jens}}}, \bibinfo {author} {\bibnamefont {{Wandelt, Benjamin D.}}},
  \bibinfo {author} {\bibnamefont {{Nissanke, Samaya}}}, \bibinfo {author}
  {\bibnamefont {{Leclercq, Florent}}}, \ and\ \bibinfo {author} {\bibnamefont
  {{Hotokezaka, Kenta}}},\ }\href {\doibase 10.1051/0004-6361/201936724}
  {\bibfield  {journal} {\bibinfo  {journal} {Astronomy \& Astrophysics}\
  }\textbf {\bibinfo {volume} {646}},\ \bibinfo {pages} {A65} (\bibinfo {year}
  {2021}{\natexlab{a}})}\BibitemShut {NoStop}%
\bibitem [{\citenamefont {Chen}(2020)}]{Chen_2020}%
  \BibitemOpen
  \bibfield  {author} {\bibinfo {author} {\bibfnamefont {H.-Y.}\ \bibnamefont
  {Chen}},\ }\href {\doibase 10.1103/PhysRevLett.125.201301} {\bibfield
  {journal} {\bibinfo  {journal} {Phys. Rev. Lett.}\ }\textbf {\bibinfo
  {volume} {125}},\ \bibinfo {pages} {201301} (\bibinfo {year}
  {2020})}\BibitemShut {NoStop}%
\bibitem [{\citenamefont {Heinzel}\ \emph {et~al.}(2021)\citenamefont
  {Heinzel}, \citenamefont {Coughlin}, \citenamefont {Dietrich}, \citenamefont
  {Bulla}, \citenamefont {Antier}, \citenamefont {Christensen}, \citenamefont
  {Coulter}, \citenamefont {Foley}, \citenamefont {Issa},\ and\ \citenamefont
  {Khetan}}]{10.1093/mnras/stab221}%
  \BibitemOpen
  \bibfield  {author} {\bibinfo {author} {\bibfnamefont {J.}~\bibnamefont
  {Heinzel}}, \bibinfo {author} {\bibfnamefont {M.~W.}\ \bibnamefont
  {Coughlin}}, \bibinfo {author} {\bibfnamefont {T.}~\bibnamefont {Dietrich}},
  \bibinfo {author} {\bibfnamefont {M.}~\bibnamefont {Bulla}}, \bibinfo
  {author} {\bibfnamefont {S.}~\bibnamefont {Antier}}, \bibinfo {author}
  {\bibfnamefont {N.}~\bibnamefont {Christensen}}, \bibinfo {author}
  {\bibfnamefont {D.~A.}\ \bibnamefont {Coulter}}, \bibinfo {author}
  {\bibfnamefont {R.~J.}\ \bibnamefont {Foley}}, \bibinfo {author}
  {\bibfnamefont {L.}~\bibnamefont {Issa}}, \ and\ \bibinfo {author}
  {\bibfnamefont {N.}~\bibnamefont {Khetan}},\ }\href {\doibase
  10.1093/mnras/stab221} {\bibfield  {journal} {\bibinfo  {journal} {Monthly
  Notices of the Royal Astronomical Society}\ }\textbf {\bibinfo {volume}
  {502}},\ \bibinfo {pages} {3057} (\bibinfo {year} {2021})},\ \Eprint
  {http://arxiv.org/abs/https://academic.oup.com/mnras/article-pdf/502/2/3057/36276043/stab221.pdf}
  {https://academic.oup.com/mnras/article-pdf/502/2/3057/36276043/stab221.pdf}
  \BibitemShut {NoStop}%
\bibitem [{\citenamefont {Sun}\ \emph {et~al.}(2020)\citenamefont {Sun},
  \citenamefont {Goetz}, \citenamefont {Kissel}, \citenamefont {Betzwieser},
  \citenamefont {Karki}, \citenamefont {Viets} \emph {et~al.}}]{Sun_2020}%
  \BibitemOpen
  \bibfield  {author} {\bibinfo {author} {\bibfnamefont {L.}~\bibnamefont
  {Sun}}, \bibinfo {author} {\bibfnamefont {E.}~\bibnamefont {Goetz}}, \bibinfo
  {author} {\bibfnamefont {J.~S.}\ \bibnamefont {Kissel}}, \bibinfo {author}
  {\bibfnamefont {J.}~\bibnamefont {Betzwieser}}, \bibinfo {author}
  {\bibfnamefont {S.}~\bibnamefont {Karki}}, \bibinfo {author} {\bibfnamefont
  {A.}~\bibnamefont {Viets}},  \emph {et~al.},\ }\href {\doibase
  10.1088/1361-6382/abb14e} {\bibfield  {journal} {\bibinfo  {journal}
  {Classical and Quantum Gravity}\ }\textbf {\bibinfo {volume} {37}},\ \bibinfo
  {pages} {225008} (\bibinfo {year} {2020})}\BibitemShut {NoStop}%
\bibitem [{\citenamefont {{The LIGO Scientific Collaboration}}\ \emph
  {et~al.}(2015)\citenamefont {{The LIGO Scientific Collaboration}} \emph
  {et~al.}}]{2015CQGra..32g4001L}%
  \BibitemOpen
  \bibfield  {author} {\bibinfo {author} {\bibnamefont {{The LIGO Scientific
  Collaboration}}} \emph {et~al.},\ }\href {\doibase
  10.1088/0264-9381/32/7/074001} {\bibfield  {journal} {\bibinfo  {journal}
  {{Class. Quant. Grav.}}\ }\textbf {\bibinfo {volume} {32}},\ \bibinfo {eid}
  {074001} (\bibinfo {year} {2015})},\ \Eprint {http://arxiv.org/abs/1411.4547}
  {arXiv:1411.4547 [gr-qc]} \BibitemShut {NoStop}%
\bibitem [{\citenamefont {Acernese}\ \emph {et~al.}(2015)\citenamefont
  {Acernese}, \citenamefont {Agathos}, \citenamefont {Agatsuma}, \citenamefont
  {Aisa}, \citenamefont {Allemandou}, \citenamefont {Allocca} \emph
  {et~al.}}]{TheVirgo:2014hva}%
  \BibitemOpen
  \bibfield  {author} {\bibinfo {author} {\bibfnamefont {F.}~\bibnamefont
  {Acernese}}, \bibinfo {author} {\bibfnamefont {M.}~\bibnamefont {Agathos}},
  \bibinfo {author} {\bibfnamefont {K.}~\bibnamefont {Agatsuma}}, \bibinfo
  {author} {\bibfnamefont {D.}~\bibnamefont {Aisa}}, \bibinfo {author}
  {\bibfnamefont {N.}~\bibnamefont {Allemandou}}, \bibinfo {author}
  {\bibfnamefont {A.}~\bibnamefont {Allocca}},  \emph {et~al.} (\bibinfo
  {collaboration} {VIRGO}),\ }\href {\doibase 10.1088/0264-9381/32/2/024001}
  {\bibfield  {journal} {\bibinfo  {journal} {Class. Quant. Grav.}\ }\textbf
  {\bibinfo {volume} {32}},\ \bibinfo {pages} {024001} (\bibinfo {year}
  {2015})},\ \Eprint {http://arxiv.org/abs/1408.3978} {arXiv:1408.3978 [gr-qc]}
  \BibitemShut {NoStop}%
\bibitem [{\citenamefont {Estevez}\ \emph {et~al.}(2021)\citenamefont
  {Estevez}, \citenamefont {Lagabbe}, \citenamefont {Masserot}, \citenamefont
  {Rolland}, \citenamefont {Seglar-Arroyo},\ and\ \citenamefont
  {Verkindt}}]{Estevez_2021}%
  \BibitemOpen
  \bibfield  {author} {\bibinfo {author} {\bibfnamefont {D.}~\bibnamefont
  {Estevez}}, \bibinfo {author} {\bibfnamefont {P.}~\bibnamefont {Lagabbe}},
  \bibinfo {author} {\bibfnamefont {A.}~\bibnamefont {Masserot}}, \bibinfo
  {author} {\bibfnamefont {L.}~\bibnamefont {Rolland}}, \bibinfo {author}
  {\bibfnamefont {M.}~\bibnamefont {Seglar-Arroyo}}, \ and\ \bibinfo {author}
  {\bibfnamefont {D.}~\bibnamefont {Verkindt}},\ }\href {\doibase
  10.1088/1361-6382/abe2db} {\bibfield  {journal} {\bibinfo  {journal}
  {Classical and Quantum Gravity}\ }\textbf {\bibinfo {volume} {38}},\ \bibinfo
  {pages} {075007} (\bibinfo {year} {2021})}\BibitemShut {NoStop}%
\bibitem [{\citenamefont {Huang}\ \emph {et~al.}(2022)\citenamefont {Huang},
  \citenamefont {Chen}, \citenamefont {Haster}, \citenamefont {Sun},
  \citenamefont {Vitale},\ and\ \citenamefont {Kissel}}]{arxiv.2204.03614}%
  \BibitemOpen
  \bibfield  {author} {\bibinfo {author} {\bibfnamefont {Y.}~\bibnamefont
  {Huang}}, \bibinfo {author} {\bibfnamefont {H.-Y.}\ \bibnamefont {Chen}},
  \bibinfo {author} {\bibfnamefont {C.-J.}\ \bibnamefont {Haster}}, \bibinfo
  {author} {\bibfnamefont {L.}~\bibnamefont {Sun}}, \bibinfo {author}
  {\bibfnamefont {S.}~\bibnamefont {Vitale}}, \ and\ \bibinfo {author}
  {\bibfnamefont {J.}~\bibnamefont {Kissel}},\ }\href {\doibase
  10.48550/ARXIV.2204.03614} {\  (\bibinfo {year} {2022}),\
  10.48550/ARXIV.2204.03614}\BibitemShut {NoStop}%
\bibitem [{\citenamefont {Veitch}\ \emph {et~al.}(2015)\citenamefont {Veitch},
  \citenamefont {Raymond}, \citenamefont {Farr}, \citenamefont {Farr},
  \citenamefont {Graff}, \citenamefont {Vitale} \emph {et~al.}}]{Veitch_2015}%
  \BibitemOpen
  \bibfield  {author} {\bibinfo {author} {\bibfnamefont {J.}~\bibnamefont
  {Veitch}}, \bibinfo {author} {\bibfnamefont {V.}~\bibnamefont {Raymond}},
  \bibinfo {author} {\bibfnamefont {B.}~\bibnamefont {Farr}}, \bibinfo {author}
  {\bibfnamefont {W.}~\bibnamefont {Farr}}, \bibinfo {author} {\bibfnamefont
  {P.}~\bibnamefont {Graff}}, \bibinfo {author} {\bibfnamefont
  {S.}~\bibnamefont {Vitale}},  \emph {et~al.},\ }\href {\doibase
  10.1103/PhysRevD.91.042003} {\bibfield  {journal} {\bibinfo  {journal} {Phys.
  Rev. D}\ }\textbf {\bibinfo {volume} {91}},\ \bibinfo {pages} {042003}
  (\bibinfo {year} {2015})}\BibitemShut {NoStop}%
\bibitem [{\citenamefont {Ashton}\ \emph {et~al.}(2019)\citenamefont {Ashton},
  \citenamefont {Hübner}, \citenamefont {Lasky}, \citenamefont {Talbot},
  \citenamefont {Ackley}, \citenamefont {Biscoveanu} \emph
  {et~al.}}]{Ashton_2019}%
  \BibitemOpen
  \bibfield  {author} {\bibinfo {author} {\bibfnamefont {G.}~\bibnamefont
  {Ashton}}, \bibinfo {author} {\bibfnamefont {M.}~\bibnamefont {Hübner}},
  \bibinfo {author} {\bibfnamefont {P.~D.}\ \bibnamefont {Lasky}}, \bibinfo
  {author} {\bibfnamefont {C.}~\bibnamefont {Talbot}}, \bibinfo {author}
  {\bibfnamefont {K.}~\bibnamefont {Ackley}}, \bibinfo {author} {\bibfnamefont
  {S.}~\bibnamefont {Biscoveanu}},  \emph {et~al.},\ }\href {\doibase
  10.3847/1538-4365/ab06fc} {\bibfield  {journal} {\bibinfo  {journal} {The
  Astrophysical Journal Supplement Series}\ }\textbf {\bibinfo {volume}
  {241}},\ \bibinfo {pages} {27} (\bibinfo {year} {2019})}\BibitemShut
  {NoStop}%
\bibitem [{\citenamefont {Romero-Shaw}\ \emph {et~al.}(2020)\citenamefont
  {Romero-Shaw}, \citenamefont {Talbot}, \citenamefont {Biscoveanu},
  \citenamefont {D’Emilio}, \citenamefont {Ashton}, \citenamefont {Berry}
  \emph {et~al.}}]{10.1093/mnras/staa2850}%
  \BibitemOpen
  \bibfield  {author} {\bibinfo {author} {\bibfnamefont {I.~M.}\ \bibnamefont
  {Romero-Shaw}}, \bibinfo {author} {\bibfnamefont {C.}~\bibnamefont {Talbot}},
  \bibinfo {author} {\bibfnamefont {S.}~\bibnamefont {Biscoveanu}}, \bibinfo
  {author} {\bibfnamefont {V.}~\bibnamefont {D’Emilio}}, \bibinfo {author}
  {\bibfnamefont {G.}~\bibnamefont {Ashton}}, \bibinfo {author} {\bibfnamefont
  {C.~P.~L.}\ \bibnamefont {Berry}},  \emph {et~al.},\ }\href {\doibase
  10.1093/mnras/staa2850} {\bibfield  {journal} {\bibinfo  {journal} {Monthly
  Notices of the Royal Astronomical Society}\ }\textbf {\bibinfo {volume}
  {499}},\ \bibinfo {pages} {3295} (\bibinfo {year} {2020})},\ \Eprint
  {http://arxiv.org/abs/https://academic.oup.com/mnras/article-pdf/499/3/3295/34052625/staa2850.pdf}
  {https://academic.oup.com/mnras/article-pdf/499/3/3295/34052625/staa2850.pdf}
  \BibitemShut {NoStop}%
\bibitem [{\citenamefont {Biwer}\ \emph {et~al.}(2019)\citenamefont {Biwer},
  \citenamefont {Capano}, \citenamefont {De}, \citenamefont {Cabero},
  \citenamefont {Brown}, \citenamefont {Nitz},\ and\ \citenamefont
  {Raymond}}]{Biwer_2019}%
  \BibitemOpen
  \bibfield  {author} {\bibinfo {author} {\bibfnamefont {C.~M.}\ \bibnamefont
  {Biwer}}, \bibinfo {author} {\bibfnamefont {C.~D.}\ \bibnamefont {Capano}},
  \bibinfo {author} {\bibfnamefont {S.}~\bibnamefont {De}}, \bibinfo {author}
  {\bibfnamefont {M.}~\bibnamefont {Cabero}}, \bibinfo {author} {\bibfnamefont
  {D.~A.}\ \bibnamefont {Brown}}, \bibinfo {author} {\bibfnamefont {A.~H.}\
  \bibnamefont {Nitz}}, \ and\ \bibinfo {author} {\bibfnamefont
  {V.}~\bibnamefont {Raymond}},\ }\href {\doibase 10.1088/1538-3873/aaef0b}
  {\bibfield  {journal} {\bibinfo  {journal} {Publications of the Astronomical
  Society of the Pacific}\ }\textbf {\bibinfo {volume} {131}},\ \bibinfo
  {pages} {024503} (\bibinfo {year} {2019})}\BibitemShut {NoStop}%
\bibitem [{\citenamefont {Lange}\ \emph {et~al.}(2018)\citenamefont {Lange},
  \citenamefont {O'Shaughnessy},\ and\ \citenamefont {Rizzo}}]{lange2018rapid}%
  \BibitemOpen
  \bibfield  {author} {\bibinfo {author} {\bibfnamefont {J.}~\bibnamefont
  {Lange}}, \bibinfo {author} {\bibfnamefont {R.}~\bibnamefont
  {O'Shaughnessy}}, \ and\ \bibinfo {author} {\bibfnamefont {M.}~\bibnamefont
  {Rizzo}},\ }\href@noop {} {\bibfield  {journal} {\bibinfo  {journal} {arXiv
  e-prints}\ } (\bibinfo {year} {2018})},\ \Eprint
  {http://arxiv.org/abs/1805.10457} {arXiv:1805.10457 [gr-qc]} \BibitemShut
  {NoStop}%
\bibitem [{\citenamefont {R\"over}(2011)}]{PhysRevD.84.122004}%
  \BibitemOpen
  \bibfield  {author} {\bibinfo {author} {\bibfnamefont {C.}~\bibnamefont
  {R\"over}},\ }\href {\doibase 10.1103/PhysRevD.84.122004} {\bibfield
  {journal} {\bibinfo  {journal} {Phys. Rev. D}\ }\textbf {\bibinfo {volume}
  {84}},\ \bibinfo {pages} {122004} (\bibinfo {year} {2011})}\BibitemShut
  {NoStop}%
\bibitem [{\citenamefont {Abbott}\ \emph
  {et~al.}(2016{\natexlab{a}})\citenamefont {Abbott} \emph
  {et~al.}}]{Abbott_2016}%
  \BibitemOpen
  \bibfield  {author} {\bibinfo {author} {\bibfnamefont {B.~P.}\ \bibnamefont
  {Abbott}} \emph {et~al.} (\bibinfo {collaboration} {The LIGO Scientific and
  Virgo Collaborations}),\ }\href {\doibase 10.1088/0264-9381/33/13/134001}
  {\bibfield  {journal} {\bibinfo  {journal} {Classical and Quantum Gravity}\
  }\textbf {\bibinfo {volume} {33}},\ \bibinfo {pages} {134001} (\bibinfo
  {year} {2016}{\natexlab{a}})}\BibitemShut {NoStop}%
\bibitem [{\citenamefont {Abbott}\ \emph
  {et~al.}(2020{\natexlab{a}})\citenamefont {Abbott} \emph
  {et~al.}}]{Abbott_2020}%
  \BibitemOpen
  \bibfield  {author} {\bibinfo {author} {\bibfnamefont {B.~P.}\ \bibnamefont
  {Abbott}} \emph {et~al.} (\bibinfo {collaboration} {The LIGO Scientific and
  Virgo Collaborations}),\ }\href {\doibase 10.1088/1361-6382/ab685e}
  {\bibfield  {journal} {\bibinfo  {journal} {Classical and Quantum Gravity}\
  }\textbf {\bibinfo {volume} {37}},\ \bibinfo {pages} {055002} (\bibinfo
  {year} {2020}{\natexlab{a}})}\BibitemShut {NoStop}%
\bibitem [{\citenamefont {Abbott}\ \emph
  {et~al.}(2021{\natexlab{a}})\citenamefont {Abbott} \emph
  {et~al.}}]{RICHABBOTT2021100658}%
  \BibitemOpen
  \bibfield  {author} {\bibinfo {author} {\bibfnamefont {R.}~\bibnamefont
  {Abbott}} \emph {et~al.} (\bibinfo {collaboration} {The LIGO Scientific and
  Virgo Collaborations}),\ }\href {\doibase
  https://doi.org/10.1016/j.softx.2021.100658} {\bibfield  {journal} {\bibinfo
  {journal} {SoftwareX}\ }\textbf {\bibinfo {volume} {13}},\ \bibinfo {pages}
  {100658} (\bibinfo {year} {2021}{\natexlab{a}})}\BibitemShut {NoStop}%
\bibitem [{\citenamefont {Davis}\ \emph {et~al.}(2021)\citenamefont {Davis}
  \emph {et~al.}}]{Davis_2021}%
  \BibitemOpen
  \bibfield  {author} {\bibinfo {author} {\bibfnamefont {D.}~\bibnamefont
  {Davis}} \emph {et~al.} (\bibinfo {collaboration} {The LIGO Scientific and
  Virgo Collaborations}),\ }\href {\doibase 10.1088/1361-6382/abfd85}
  {\bibfield  {journal} {\bibinfo  {journal} {Classical and Quantum Gravity}\
  }\textbf {\bibinfo {volume} {38}},\ \bibinfo {pages} {135014} (\bibinfo
  {year} {2021})}\BibitemShut {NoStop}%
\bibitem [{\citenamefont {Nuttall}\ \emph {et~al.}(2015)\citenamefont
  {Nuttall}, \citenamefont {Massinger}, \citenamefont {Areeda}, \citenamefont
  {Betzwieser}, \citenamefont {Dwyer}, \citenamefont {Effler} \emph
  {et~al.}}]{Nuttall_2015}%
  \BibitemOpen
  \bibfield  {author} {\bibinfo {author} {\bibfnamefont {L.~K.}\ \bibnamefont
  {Nuttall}}, \bibinfo {author} {\bibfnamefont {T.~J.}\ \bibnamefont
  {Massinger}}, \bibinfo {author} {\bibfnamefont {J.}~\bibnamefont {Areeda}},
  \bibinfo {author} {\bibfnamefont {J.}~\bibnamefont {Betzwieser}}, \bibinfo
  {author} {\bibfnamefont {S.}~\bibnamefont {Dwyer}}, \bibinfo {author}
  {\bibfnamefont {A.}~\bibnamefont {Effler}},  \emph {et~al.},\ }\href
  {\doibase 10.1088/0264-9381/32/24/245005} {\bibfield  {journal} {\bibinfo
  {journal} {Classical and Quantum Gravity}\ }\textbf {\bibinfo {volume}
  {32}},\ \bibinfo {pages} {245005} (\bibinfo {year} {2015})}\BibitemShut
  {NoStop}%
\bibitem [{\citenamefont {Zevin}\ \emph {et~al.}(2017)\citenamefont {Zevin},
  \citenamefont {Coughlin}, \citenamefont {Bahaadini}, \citenamefont {Besler},
  \citenamefont {Rohani}, \citenamefont {Allen} \emph {et~al.}}]{Zevin_2017}%
  \BibitemOpen
  \bibfield  {author} {\bibinfo {author} {\bibfnamefont {M.}~\bibnamefont
  {Zevin}}, \bibinfo {author} {\bibfnamefont {S.}~\bibnamefont {Coughlin}},
  \bibinfo {author} {\bibfnamefont {S.}~\bibnamefont {Bahaadini}}, \bibinfo
  {author} {\bibfnamefont {E.}~\bibnamefont {Besler}}, \bibinfo {author}
  {\bibfnamefont {N.}~\bibnamefont {Rohani}}, \bibinfo {author} {\bibfnamefont
  {S.}~\bibnamefont {Allen}},  \emph {et~al.},\ }\href {\doibase
  10.1088/1361-6382/aa5cea} {\bibfield  {journal} {\bibinfo  {journal}
  {Classical and Quantum Gravity}\ }\textbf {\bibinfo {volume} {34}},\ \bibinfo
  {pages} {064003} (\bibinfo {year} {2017})}\BibitemShut {NoStop}%
\bibitem [{\citenamefont {Abbott}\ \emph
  {et~al.}(2017{\natexlab{b}})\citenamefont {Abbott} \emph
  {et~al.}}]{LIGOScientific:2017vwq}%
  \BibitemOpen
  \bibfield  {author} {\bibinfo {author} {\bibfnamefont {B.~P.}\ \bibnamefont
  {Abbott}} \emph {et~al.} (\bibinfo {collaboration} {LIGO Scientific,
  Virgo}),\ }\href {\doibase 10.1103/PhysRevLett.119.161101} {\bibfield
  {journal} {\bibinfo  {journal} {Phys. Rev. Lett.}\ }\textbf {\bibinfo
  {volume} {119}},\ \bibinfo {pages} {161101} (\bibinfo {year}
  {2017}{\natexlab{b}})},\ \Eprint {http://arxiv.org/abs/1710.05832}
  {arXiv:1710.05832 [gr-qc]} \BibitemShut {NoStop}%
\bibitem [{\citenamefont {Pankow}\ \emph {et~al.}(2018)\citenamefont {Pankow},
  \citenamefont {Chatziioannou}, \citenamefont {Chase}, \citenamefont
  {Littenberg}, \citenamefont {Evans}, \citenamefont {McIver} \emph
  {et~al.}}]{Pankow_2018}%
  \BibitemOpen
  \bibfield  {author} {\bibinfo {author} {\bibfnamefont {C.}~\bibnamefont
  {Pankow}}, \bibinfo {author} {\bibfnamefont {K.}~\bibnamefont
  {Chatziioannou}}, \bibinfo {author} {\bibfnamefont {E.~A.}\ \bibnamefont
  {Chase}}, \bibinfo {author} {\bibfnamefont {T.~B.}\ \bibnamefont
  {Littenberg}}, \bibinfo {author} {\bibfnamefont {M.}~\bibnamefont {Evans}},
  \bibinfo {author} {\bibfnamefont {J.}~\bibnamefont {McIver}},  \emph
  {et~al.},\ }\href {\doibase 10.1103/PhysRevD.98.084016} {\bibfield  {journal}
  {\bibinfo  {journal} {Phys. Rev. D}\ }\textbf {\bibinfo {volume} {98}},\
  \bibinfo {pages} {084016} (\bibinfo {year} {2018})}\BibitemShut {NoStop}%
\bibitem [{\citenamefont {Abbott}\ \emph
  {et~al.}(2021{\natexlab{b}})\citenamefont {Abbott} \emph
  {et~al.}}]{LIGOScientific:2020ibl}%
  \BibitemOpen
  \bibfield  {author} {\bibinfo {author} {\bibfnamefont {R.}~\bibnamefont
  {Abbott}} \emph {et~al.} (\bibinfo {collaboration} {LIGO Scientific,
  Virgo}),\ }\href {\doibase 10.1103/PhysRevX.11.021053} {\bibfield  {journal}
  {\bibinfo  {journal} {Phys. Rev. X}\ }\textbf {\bibinfo {volume} {11}},\
  \bibinfo {pages} {021053} (\bibinfo {year} {2021}{\natexlab{b}})},\ \Eprint
  {http://arxiv.org/abs/2010.14527} {arXiv:2010.14527 [gr-qc]} \BibitemShut
  {NoStop}%
\bibitem [{\citenamefont {Vallisneri}(2008)}]{PhysRevD.77.042001}%
  \BibitemOpen
  \bibfield  {author} {\bibinfo {author} {\bibfnamefont {M.}~\bibnamefont
  {Vallisneri}},\ }\href {\doibase 10.1103/PhysRevD.77.042001} {\bibfield
  {journal} {\bibinfo  {journal} {Phys. Rev. D}\ }\textbf {\bibinfo {volume}
  {77}},\ \bibinfo {pages} {042001} (\bibinfo {year} {2008})}\BibitemShut
  {NoStop}%
\bibitem [{\citenamefont {Edy}\ \emph {et~al.}(2021)\citenamefont {Edy},
  \citenamefont {Lundgren},\ and\ \citenamefont {Nuttall}}]{Edy_2021}%
  \BibitemOpen
  \bibfield  {author} {\bibinfo {author} {\bibfnamefont {O.}~\bibnamefont
  {Edy}}, \bibinfo {author} {\bibfnamefont {A.}~\bibnamefont {Lundgren}}, \
  and\ \bibinfo {author} {\bibfnamefont {L.~K.}\ \bibnamefont {Nuttall}},\
  }\href {\doibase 10.1103/PhysRevD.103.124061} {\bibfield  {journal} {\bibinfo
   {journal} {Phys. Rev. D}\ }\textbf {\bibinfo {volume} {103}},\ \bibinfo
  {pages} {124061} (\bibinfo {year} {2021})}\BibitemShut {NoStop}%
\bibitem [{\citenamefont {Vallisneri}\ \emph {et~al.}(2015)\citenamefont
  {Vallisneri}, \citenamefont {Kanner}, \citenamefont {Williams}, \citenamefont
  {Weinstein},\ and\ \citenamefont {Stephens}}]{Vallisneri_2015}%
  \BibitemOpen
  \bibfield  {author} {\bibinfo {author} {\bibfnamefont {M.}~\bibnamefont
  {Vallisneri}}, \bibinfo {author} {\bibfnamefont {J.}~\bibnamefont {Kanner}},
  \bibinfo {author} {\bibfnamefont {R.}~\bibnamefont {Williams}}, \bibinfo
  {author} {\bibfnamefont {A.}~\bibnamefont {Weinstein}}, \ and\ \bibinfo
  {author} {\bibfnamefont {B.}~\bibnamefont {Stephens}},\ }\href {\doibase
  10.1088/1742-6596/610/1/012021} {\bibfield  {journal} {\bibinfo  {journal}
  {Journal of Physics: Conference Series}\ }\textbf {\bibinfo {volume} {610}},\
  \bibinfo {pages} {012021} (\bibinfo {year} {2015})}\BibitemShut {NoStop}%
\bibitem [{\citenamefont {Cornish}(2020)}]{PhysRevD.102.124038}%
  \BibitemOpen
  \bibfield  {author} {\bibinfo {author} {\bibfnamefont {N.~J.}\ \bibnamefont
  {Cornish}},\ }\href {\doibase 10.1103/PhysRevD.102.124038} {\bibfield
  {journal} {\bibinfo  {journal} {Phys. Rev. D}\ }\textbf {\bibinfo {volume}
  {102}},\ \bibinfo {pages} {124038} (\bibinfo {year} {2020})}\BibitemShut
  {NoStop}%
\bibitem [{\citenamefont {Cornish}\ \emph {et~al.}(2021)\citenamefont
  {Cornish}, \citenamefont {Littenberg}, \citenamefont {B\'ecsy}, \citenamefont
  {Chatziioannou}, \citenamefont {Clark}, \citenamefont {Ghonge},\ and\
  \citenamefont {Millhouse}}]{PhysRevD.103.044006}%
  \BibitemOpen
  \bibfield  {author} {\bibinfo {author} {\bibfnamefont {N.~J.}\ \bibnamefont
  {Cornish}}, \bibinfo {author} {\bibfnamefont {T.~B.}\ \bibnamefont
  {Littenberg}}, \bibinfo {author} {\bibfnamefont {B.}~\bibnamefont {B\'ecsy}},
  \bibinfo {author} {\bibfnamefont {K.}~\bibnamefont {Chatziioannou}}, \bibinfo
  {author} {\bibfnamefont {J.~A.}\ \bibnamefont {Clark}}, \bibinfo {author}
  {\bibfnamefont {S.}~\bibnamefont {Ghonge}}, \ and\ \bibinfo {author}
  {\bibfnamefont {M.}~\bibnamefont {Millhouse}},\ }\href {\doibase
  10.1103/PhysRevD.103.044006} {\bibfield  {journal} {\bibinfo  {journal}
  {Phys. Rev. D}\ }\textbf {\bibinfo {volume} {103}},\ \bibinfo {pages}
  {044006} (\bibinfo {year} {2021})}\BibitemShut {NoStop}%
\bibitem [{\citenamefont {Cutler}\ \emph {et~al.}(1993)\citenamefont {Cutler},
  \citenamefont {Apostolatos}, \citenamefont {Bildsten}, \citenamefont {Finn},
  \citenamefont {Flanagan}, \citenamefont {Kennefick} \emph
  {et~al.}}]{Cutler_1993}%
  \BibitemOpen
  \bibfield  {author} {\bibinfo {author} {\bibfnamefont {C.}~\bibnamefont
  {Cutler}}, \bibinfo {author} {\bibfnamefont {T.~A.}\ \bibnamefont
  {Apostolatos}}, \bibinfo {author} {\bibfnamefont {L.}~\bibnamefont
  {Bildsten}}, \bibinfo {author} {\bibfnamefont {L.~S.}\ \bibnamefont {Finn}},
  \bibinfo {author} {\bibfnamefont {E.~E.}\ \bibnamefont {Flanagan}}, \bibinfo
  {author} {\bibfnamefont {D.}~\bibnamefont {Kennefick}},  \emph {et~al.},\
  }\href {\doibase 10.1103/PhysRevLett.70.2984} {\bibfield  {journal} {\bibinfo
   {journal} {Phys. Rev. Lett.}\ }\textbf {\bibinfo {volume} {70}},\ \bibinfo
  {pages} {2984} (\bibinfo {year} {1993})}\BibitemShut {NoStop}%
\bibitem [{\citenamefont {Allen}\ \emph {et~al.}(2012)\citenamefont {Allen},
  \citenamefont {Anderson}, \citenamefont {Brady}, \citenamefont {Brown},\ and\
  \citenamefont {Creighton}}]{Allen_2012}%
  \BibitemOpen
  \bibfield  {author} {\bibinfo {author} {\bibfnamefont {B.}~\bibnamefont
  {Allen}}, \bibinfo {author} {\bibfnamefont {W.~G.}\ \bibnamefont {Anderson}},
  \bibinfo {author} {\bibfnamefont {P.~R.}\ \bibnamefont {Brady}}, \bibinfo
  {author} {\bibfnamefont {D.~A.}\ \bibnamefont {Brown}}, \ and\ \bibinfo
  {author} {\bibfnamefont {J.~D.~E.}\ \bibnamefont {Creighton}},\ }\href
  {\doibase 10.1103/PhysRevD.85.122006} {\bibfield  {journal} {\bibinfo
  {journal} {Phys. Rev. D}\ }\textbf {\bibinfo {volume} {85}},\ \bibinfo
  {pages} {122006} (\bibinfo {year} {2012})}\BibitemShut {NoStop}%
\bibitem [{\citenamefont {Finn}(1992)}]{PhysRevD.46.5236}%
  \BibitemOpen
  \bibfield  {author} {\bibinfo {author} {\bibfnamefont {L.~S.}\ \bibnamefont
  {Finn}},\ }\href {\doibase 10.1103/PhysRevD.46.5236} {\bibfield  {journal}
  {\bibinfo  {journal} {Phys. Rev. D}\ }\textbf {\bibinfo {volume} {46}},\
  \bibinfo {pages} {5236} (\bibinfo {year} {1992})}\BibitemShut {NoStop}%
\bibitem [{\citenamefont {Romano}\ and\ \citenamefont
  {Cornish}(2017)}]{Romano_2017}%
  \BibitemOpen
  \bibfield  {author} {\bibinfo {author} {\bibfnamefont {J.~D.}\ \bibnamefont
  {Romano}}\ and\ \bibinfo {author} {\bibfnamefont {N.~J.}\ \bibnamefont
  {Cornish}},\ }\href {\doibase 10.1007/s41114-017-0004-1} {\bibfield
  {journal} {\bibinfo  {journal} {Living Reviews in Relativity}\ }\textbf
  {\bibinfo {volume} {20}} (\bibinfo {year} {2017}),\
  10.1007/s41114-017-0004-1}\BibitemShut {NoStop}%
\bibitem [{\citenamefont {Veitch}\ and\ \citenamefont
  {Vecchio}(2010)}]{Veitch_2010}%
  \BibitemOpen
  \bibfield  {author} {\bibinfo {author} {\bibfnamefont {J.}~\bibnamefont
  {Veitch}}\ and\ \bibinfo {author} {\bibfnamefont {A.}~\bibnamefont
  {Vecchio}},\ }\href {\doibase 10.1103/PhysRevD.81.062003} {\bibfield
  {journal} {\bibinfo  {journal} {Phys. Rev. D}\ }\textbf {\bibinfo {volume}
  {81}},\ \bibinfo {pages} {062003} (\bibinfo {year} {2010})}\BibitemShut
  {NoStop}%
\bibitem [{\citenamefont {Littenberg}\ and\ \citenamefont
  {Cornish}(2015)}]{PhysRevD.91.084034}%
  \BibitemOpen
  \bibfield  {author} {\bibinfo {author} {\bibfnamefont {T.~B.}\ \bibnamefont
  {Littenberg}}\ and\ \bibinfo {author} {\bibfnamefont {N.~J.}\ \bibnamefont
  {Cornish}},\ }\href {\doibase 10.1103/PhysRevD.91.084034} {\bibfield
  {journal} {\bibinfo  {journal} {Phys. Rev. D}\ }\textbf {\bibinfo {volume}
  {91}},\ \bibinfo {pages} {084034} (\bibinfo {year} {2015})}\BibitemShut
  {NoStop}%
\bibitem [{\citenamefont {Talbot}\ \emph {et~al.}(2021)\citenamefont {Talbot},
  \citenamefont {Thrane}, \citenamefont {Biscoveanu},\ and\ \citenamefont
  {Smith}}]{talbot2021inference}%
  \BibitemOpen
  \bibfield  {author} {\bibinfo {author} {\bibfnamefont {C.}~\bibnamefont
  {Talbot}}, \bibinfo {author} {\bibfnamefont {E.}~\bibnamefont {Thrane}},
  \bibinfo {author} {\bibfnamefont {S.}~\bibnamefont {Biscoveanu}}, \ and\
  \bibinfo {author} {\bibfnamefont {R.}~\bibnamefont {Smith}},\ }\href
  {\doibase 10.1103/PhysRevResearch.3.043049} {\bibfield  {journal} {\bibinfo
  {journal} {Phys. Rev. Res.}\ }\textbf {\bibinfo {volume} {3}},\ \bibinfo
  {pages} {043049} (\bibinfo {year} {2021})},\ \Eprint
  {http://arxiv.org/abs/2106.13785} {arXiv:2106.13785 [astro-ph.IM]}
  \BibitemShut {NoStop}%
\bibitem [{\citenamefont {{Talbot}}\ and\ \citenamefont
  {{Thrane}}(2020)}]{2020PhRvR...2d3298T}%
  \BibitemOpen
  \bibfield  {author} {\bibinfo {author} {\bibfnamefont {C.}~\bibnamefont
  {{Talbot}}}\ and\ \bibinfo {author} {\bibfnamefont {E.}~\bibnamefont
  {{Thrane}}},\ }\href {\doibase 10.1103/PhysRevResearch.2.043298} {\bibfield
  {journal} {\bibinfo  {journal} {Physical Review Research}\ }\textbf {\bibinfo
  {volume} {2}},\ \bibinfo {eid} {043298} (\bibinfo {year} {2020})},\ \Eprint
  {http://arxiv.org/abs/2006.05292} {arXiv:2006.05292 [astro-ph.IM]}
  \BibitemShut {NoStop}%
\bibitem [{\citenamefont {Chatziioannou}\ \emph {et~al.}(2019)\citenamefont
  {Chatziioannou}, \citenamefont {Haster}, \citenamefont {Littenberg},
  \citenamefont {Farr}, \citenamefont {Ghonge}, \citenamefont {Millhouse},
  \citenamefont {Clark},\ and\ \citenamefont {Cornish}}]{Chatziioannou_2019}%
  \BibitemOpen
  \bibfield  {author} {\bibinfo {author} {\bibfnamefont {K.}~\bibnamefont
  {Chatziioannou}}, \bibinfo {author} {\bibfnamefont {C.-J.}\ \bibnamefont
  {Haster}}, \bibinfo {author} {\bibfnamefont {T.~B.}\ \bibnamefont
  {Littenberg}}, \bibinfo {author} {\bibfnamefont {W.~M.}\ \bibnamefont
  {Farr}}, \bibinfo {author} {\bibfnamefont {S.}~\bibnamefont {Ghonge}},
  \bibinfo {author} {\bibfnamefont {M.}~\bibnamefont {Millhouse}}, \bibinfo
  {author} {\bibfnamefont {J.~A.}\ \bibnamefont {Clark}}, \ and\ \bibinfo
  {author} {\bibfnamefont {N.}~\bibnamefont {Cornish}},\ }\href {\doibase
  10.1103/PhysRevD.100.104004} {\bibfield  {journal} {\bibinfo  {journal}
  {Phys. Rev. D}\ }\textbf {\bibinfo {volume} {100}},\ \bibinfo {pages}
  {104004} (\bibinfo {year} {2019})}\BibitemShut {NoStop}%
\bibitem [{\citenamefont {Cornish}\ and\ \citenamefont
  {Littenberg}(2015)}]{Cornish_2015}%
  \BibitemOpen
  \bibfield  {author} {\bibinfo {author} {\bibfnamefont {N.~J.}\ \bibnamefont
  {Cornish}}\ and\ \bibinfo {author} {\bibfnamefont {T.~B.}\ \bibnamefont
  {Littenberg}},\ }\href {\doibase 10.1088/0264-9381/32/13/135012} {\bibfield
  {journal} {\bibinfo  {journal} {Classical and Quantum Gravity}\ }\textbf
  {\bibinfo {volume} {32}},\ \bibinfo {pages} {135012} (\bibinfo {year}
  {2015})}\BibitemShut {NoStop}%
\bibitem [{\citenamefont {Biscoveanu}\ \emph {et~al.}(2020)\citenamefont
  {Biscoveanu}, \citenamefont {Haster}, \citenamefont {Vitale},\ and\
  \citenamefont {Davies}}]{Biscoveanu_2020}%
  \BibitemOpen
  \bibfield  {author} {\bibinfo {author} {\bibfnamefont {S.}~\bibnamefont
  {Biscoveanu}}, \bibinfo {author} {\bibfnamefont {C.-J.}\ \bibnamefont
  {Haster}}, \bibinfo {author} {\bibfnamefont {S.}~\bibnamefont {Vitale}}, \
  and\ \bibinfo {author} {\bibfnamefont {J.}~\bibnamefont {Davies}},\ }\href
  {\doibase 10.1103/PhysRevD.102.023008} {\bibfield  {journal} {\bibinfo
  {journal} {Phys. Rev. D}\ }\textbf {\bibinfo {volume} {102}},\ \bibinfo
  {pages} {023008} (\bibinfo {year} {2020})}\BibitemShut {NoStop}%
\bibitem [{\citenamefont {Mozzon}\ \emph {et~al.}(2020)\citenamefont {Mozzon},
  , \citenamefont {Nuttall}, \citenamefont {Lundgren}, \citenamefont {Dent},
  \citenamefont {Kumar},\ and\ \citenamefont {Nitz}}]{Mozzon_2020}%
  \BibitemOpen
  \bibfield  {author} {\bibinfo {author} {\bibfnamefont {S.}~\bibnamefont
  {Mozzon}}, , \bibinfo {author} {\bibfnamefont {L.~K.}\ \bibnamefont
  {Nuttall}}, \bibinfo {author} {\bibfnamefont {A.}~\bibnamefont {Lundgren}},
  \bibinfo {author} {\bibfnamefont {T.}~\bibnamefont {Dent}}, \bibinfo {author}
  {\bibfnamefont {S.}~\bibnamefont {Kumar}}, \ and\ \bibinfo {author}
  {\bibfnamefont {A.~H.}\ \bibnamefont {Nitz}},\ }\href {\doibase
  10.1088/1361-6382/abac6c} {\bibfield  {journal} {\bibinfo  {journal}
  {Classical and Quantum Gravity}\ }\textbf {\bibinfo {volume} {37}},\ \bibinfo
  {pages} {215014} (\bibinfo {year} {2020})}\BibitemShut {NoStop}%
\bibitem [{\citenamefont {Abbott}\ \emph
  {et~al.}(2020{\natexlab{b}})\citenamefont {Abbott} \emph
  {et~al.}}]{Abbott_2020_1}%
  \BibitemOpen
  \bibfield  {author} {\bibinfo {author} {\bibfnamefont {B.~P.}\ \bibnamefont
  {Abbott}} \emph {et~al.},\ }\href {\doibase 10.1007/s41114-020-00026-9}
  {\bibfield  {journal} {\bibinfo  {journal} {Living Reviews in Relativity}\
  }\textbf {\bibinfo {volume} {23}} (\bibinfo {year} {2020}{\natexlab{b}}),\
  10.1007/s41114-020-00026-9}\BibitemShut {NoStop}%
\bibitem [{\citenamefont {Abbott}\ \emph
  {et~al.}(2021{\natexlab{c}})\citenamefont {Abbott} \emph
  {et~al.}}]{Abbott:2019yzh}%
  \BibitemOpen
  \bibfield  {author} {\bibinfo {author} {\bibfnamefont {B.~P.}\ \bibnamefont
  {Abbott}} \emph {et~al.} (\bibinfo {collaboration} {LIGO Scientific,
  Virgo}),\ }\href {\doibase 10.3847/1538-4357/abdcb7} {\bibfield  {journal}
  {\bibinfo  {journal} {Astrophys. J.}\ }\textbf {\bibinfo {volume} {909}},\
  \bibinfo {pages} {218} (\bibinfo {year} {2021}{\natexlab{c}})},\ \Eprint
  {http://arxiv.org/abs/1908.06060} {arXiv:1908.06060 [astro-ph.CO]}
  \BibitemShut {NoStop}%
\bibitem [{\citenamefont {Abbott}\ \emph
  {et~al.}(2016{\natexlab{b}})\citenamefont {Abbott} \emph
  {et~al.}}]{TheLIGOScientific:2016zmo}%
  \BibitemOpen
  \bibfield  {author} {\bibinfo {author} {\bibfnamefont {B.~P.}\ \bibnamefont
  {Abbott}} \emph {et~al.} (\bibinfo {collaboration} {LIGO Scientific,
  Virgo}),\ }\href {\doibase 10.1088/0264-9381/33/13/134001} {\bibfield
  {journal} {\bibinfo  {journal} {Class. Quant. Grav.}\ }\textbf {\bibinfo
  {volume} {33}},\ \bibinfo {pages} {134001} (\bibinfo {year}
  {2016}{\natexlab{b}})},\ \Eprint {http://arxiv.org/abs/1602.03844}
  {arXiv:1602.03844 [gr-qc]} \BibitemShut {NoStop}%
\bibitem [{\citenamefont {{Nuttall}}(2018)}]{2018RSPTA.37670286N}%
  \BibitemOpen
  \bibfield  {author} {\bibinfo {author} {\bibfnamefont {L.~K.}\ \bibnamefont
  {{Nuttall}}},\ }\href {\doibase 10.1098/rsta.2017.0286} {\bibfield  {journal}
  {\bibinfo  {journal} {Philosophical Transactions of the Royal Society of
  London Series A}\ }\textbf {\bibinfo {volume} {376}},\ \bibinfo {eid}
  {20170286} (\bibinfo {year} {2018})},\ \Eprint
  {http://arxiv.org/abs/1804.07592} {arXiv:1804.07592 [astro-ph.IM]}
  \BibitemShut {NoStop}%
\bibitem [{\citenamefont {Sathyaprakash}\ and\ \citenamefont
  {Schutz}(2009)}]{Sathyaprakash_2009}%
  \BibitemOpen
  \bibfield  {author} {\bibinfo {author} {\bibfnamefont {B.~S.}\ \bibnamefont
  {Sathyaprakash}}\ and\ \bibinfo {author} {\bibfnamefont {B.~F.}\ \bibnamefont
  {Schutz}},\ }\href {\doibase 10.12942/lrr-2009-2} {\bibfield  {journal}
  {\bibinfo  {journal} {Living Reviews in Relativity}\ }\textbf {\bibinfo
  {volume} {12}} (\bibinfo {year} {2009}),\ 10.12942/lrr-2009-2}\BibitemShut
  {NoStop}%
\bibitem [{\citenamefont {Abbott}\ \emph
  {et~al.}(2016{\natexlab{c}})\citenamefont {Abbott} \emph
  {et~al.}}]{PhysRevLett.116.241102}%
  \BibitemOpen
  \bibfield  {author} {\bibinfo {author} {\bibfnamefont {B.~P.}\ \bibnamefont
  {Abbott}} \emph {et~al.} (\bibinfo {collaboration} {LIGO Scientific
  Collaboration and Virgo Collaboration}),\ }\href {\doibase
  10.1103/PhysRevLett.116.241102} {\bibfield  {journal} {\bibinfo  {journal}
  {Phys. Rev. Lett.}\ }\textbf {\bibinfo {volume} {116}},\ \bibinfo {pages}
  {241102} (\bibinfo {year} {2016}{\natexlab{c}})}\BibitemShut {NoStop}%
\bibitem [{\citenamefont {Abbott}\ \emph {et~al.}(2019)\citenamefont {Abbott}
  \emph {et~al.}}]{PhysRevX.9.031040}%
  \BibitemOpen
  \bibfield  {author} {\bibinfo {author} {\bibfnamefont {B.~P.}\ \bibnamefont
  {Abbott}} \emph {et~al.} (\bibinfo {collaboration} {LIGO Scientific
  Collaboration and Virgo Collaboration}),\ }\href {\doibase
  10.1103/PhysRevX.9.031040} {\bibfield  {journal} {\bibinfo  {journal} {Phys.
  Rev. X}\ }\textbf {\bibinfo {volume} {9}},\ \bibinfo {pages} {031040}
  (\bibinfo {year} {2019})}\BibitemShut {NoStop}%
\bibitem [{\citenamefont {Zhu}\ \emph {et~al.}(2018)\citenamefont {Zhu},
  \citenamefont {Thrane}, \citenamefont {Os\l{}owski}, \citenamefont {Levin},\
  and\ \citenamefont {Lasky}}]{Zhu:2017znf}%
  \BibitemOpen
  \bibfield  {author} {\bibinfo {author} {\bibfnamefont {X.}~\bibnamefont
  {Zhu}}, \bibinfo {author} {\bibfnamefont {E.}~\bibnamefont {Thrane}},
  \bibinfo {author} {\bibfnamefont {S.}~\bibnamefont {Os\l{}owski}}, \bibinfo
  {author} {\bibfnamefont {Y.}~\bibnamefont {Levin}}, \ and\ \bibinfo {author}
  {\bibfnamefont {P.~D.}\ \bibnamefont {Lasky}},\ }\href {\doibase
  10.1103/PhysRevD.98.043002} {\bibfield  {journal} {\bibinfo  {journal} {Phys.
  Rev. D}\ }\textbf {\bibinfo {volume} {98}},\ \bibinfo {pages} {043002}
  (\bibinfo {year} {2018})}\BibitemShut {NoStop}%
\bibitem [{\citenamefont {Abbott}\ \emph
  {et~al.}(2021{\natexlab{d}})\citenamefont {Abbott} \emph
  {et~al.}}]{PhysRevX.11.021053}%
  \BibitemOpen
  \bibfield  {author} {\bibinfo {author} {\bibfnamefont {R.}~\bibnamefont
  {Abbott}} \emph {et~al.} (\bibinfo {collaboration} {The LIGO Scientific and
  Virgo Collaborations}),\ }\href {\doibase 10.1103/PhysRevX.11.021053}
  {\bibfield  {journal} {\bibinfo  {journal} {Phys. Rev. X}\ }\textbf {\bibinfo
  {volume} {11}},\ \bibinfo {pages} {021053} (\bibinfo {year}
  {2021}{\natexlab{d}})}\BibitemShut {NoStop}%
\bibitem [{\citenamefont {Schmidt}\ \emph {et~al.}(2015)\citenamefont
  {Schmidt}, \citenamefont {Ohme},\ and\ \citenamefont
  {Hannam}}]{PhysRevD.91.024043}%
  \BibitemOpen
  \bibfield  {author} {\bibinfo {author} {\bibfnamefont {P.}~\bibnamefont
  {Schmidt}}, \bibinfo {author} {\bibfnamefont {F.}~\bibnamefont {Ohme}}, \
  and\ \bibinfo {author} {\bibfnamefont {M.}~\bibnamefont {Hannam}},\ }\href
  {\doibase 10.1103/PhysRevD.91.024043} {\bibfield  {journal} {\bibinfo
  {journal} {Phys. Rev. D}\ }\textbf {\bibinfo {volume} {91}},\ \bibinfo
  {pages} {024043} (\bibinfo {year} {2015})}\BibitemShut {NoStop}%
\bibitem [{\citenamefont {Hannam}\ \emph {et~al.}(2014)\citenamefont {Hannam},
  \citenamefont {Schmidt}, \citenamefont {Boh\'e}, \citenamefont {Haegel},
  \citenamefont {Husa}, \citenamefont {Ohme}, \citenamefont {Pratten},\ and\
  \citenamefont {P\"urrer}}]{PhysRevLett.113.151101}%
  \BibitemOpen
  \bibfield  {author} {\bibinfo {author} {\bibfnamefont {M.}~\bibnamefont
  {Hannam}}, \bibinfo {author} {\bibfnamefont {P.}~\bibnamefont {Schmidt}},
  \bibinfo {author} {\bibfnamefont {A.}~\bibnamefont {Boh\'e}}, \bibinfo
  {author} {\bibfnamefont {L.}~\bibnamefont {Haegel}}, \bibinfo {author}
  {\bibfnamefont {S.}~\bibnamefont {Husa}}, \bibinfo {author} {\bibfnamefont
  {F.}~\bibnamefont {Ohme}}, \bibinfo {author} {\bibfnamefont {G.}~\bibnamefont
  {Pratten}}, \ and\ \bibinfo {author} {\bibfnamefont {M.}~\bibnamefont
  {P\"urrer}},\ }\href {\doibase 10.1103/PhysRevLett.113.151101} {\bibfield
  {journal} {\bibinfo  {journal} {Phys. Rev. Lett.}\ }\textbf {\bibinfo
  {volume} {113}},\ \bibinfo {pages} {151101} (\bibinfo {year}
  {2014})}\BibitemShut {NoStop}%
\bibitem [{\citenamefont {Smith}\ \emph {et~al.}(2021)\citenamefont {Smith},
  \citenamefont {Borhanian}, \citenamefont {Sathyaprakash}, \citenamefont
  {Hernandez~Vivanco}, \citenamefont {Field}, \citenamefont {Lasky} \emph
  {et~al.}}]{pizzati2021bayesian}%
  \BibitemOpen
  \bibfield  {author} {\bibinfo {author} {\bibfnamefont {R.}~\bibnamefont
  {Smith}}, \bibinfo {author} {\bibfnamefont {S.}~\bibnamefont {Borhanian}},
  \bibinfo {author} {\bibfnamefont {B.}~\bibnamefont {Sathyaprakash}}, \bibinfo
  {author} {\bibfnamefont {F.}~\bibnamefont {Hernandez~Vivanco}}, \bibinfo
  {author} {\bibfnamefont {S.~E.}\ \bibnamefont {Field}}, \bibinfo {author}
  {\bibfnamefont {P.}~\bibnamefont {Lasky}},  \emph {et~al.},\ }\href {\doibase
  10.1103/PhysRevLett.127.081102} {\bibfield  {journal} {\bibinfo  {journal}
  {Phys. Rev. Lett.}\ }\textbf {\bibinfo {volume} {127}},\ \bibinfo {pages}
  {081102} (\bibinfo {year} {2021})}\BibitemShut {NoStop}%
\bibitem [{\citenamefont {Samajdar}\ \emph {et~al.}(2021)\citenamefont
  {Samajdar}, \citenamefont {Janquart}, \citenamefont {VanDenBroeck},\ and\
  \citenamefont {Dietrich}}]{Samajdar_2021}%
  \BibitemOpen
  \bibfield  {author} {\bibinfo {author} {\bibfnamefont {A.}~\bibnamefont
  {Samajdar}}, \bibinfo {author} {\bibfnamefont {J.}~\bibnamefont {Janquart}},
  \bibinfo {author} {\bibfnamefont {C.}~\bibnamefont {VanDenBroeck}}, \ and\
  \bibinfo {author} {\bibfnamefont {T.}~\bibnamefont {Dietrich}},\ }\href
  {\doibase 10.1103/PhysRevD.104.044003} {\bibfield  {journal} {\bibinfo
  {journal} {Phys. Rev. D}\ }\textbf {\bibinfo {volume} {104}},\ \bibinfo
  {pages} {044003} (\bibinfo {year} {2021})}\BibitemShut {NoStop}%
\bibitem [{\citenamefont {Relton}\ and\ \citenamefont
  {Raymond}(2021)}]{relton2021parameter}%
  \BibitemOpen
  \bibfield  {author} {\bibinfo {author} {\bibfnamefont {P.}~\bibnamefont
  {Relton}}\ and\ \bibinfo {author} {\bibfnamefont {V.}~\bibnamefont
  {Raymond}},\ }\href {\doibase 10.1103/PhysRevD.104.084039} {\bibfield
  {journal} {\bibinfo  {journal} {Phys. Rev. D}\ }\textbf {\bibinfo {volume}
  {104}},\ \bibinfo {pages} {084039} (\bibinfo {year} {2021})}\BibitemShut
  {NoStop}%
\bibitem [{\citenamefont {{Speagle}}(2020)}]{2020MNRAS.493.3132S}%
  \BibitemOpen
  \bibfield  {author} {\bibinfo {author} {\bibfnamefont {J.~S.}\ \bibnamefont
  {{Speagle}}},\ }\href {\doibase 10.1093/mnras/staa278} {\bibfield  {journal}
  {\bibinfo  {journal} {\mnras}\ }\textbf {\bibinfo {volume} {493}},\ \bibinfo
  {pages} {3132} (\bibinfo {year} {2020})},\ \Eprint
  {http://arxiv.org/abs/1904.02180} {arXiv:1904.02180 [astro-ph.IM]}
  \BibitemShut {NoStop}%
\bibitem [{\citenamefont {Canizares}\ \emph {et~al.}(2015)\citenamefont
  {Canizares}, \citenamefont {Field}, \citenamefont {Gair}, \citenamefont
  {Raymond}, \citenamefont {Smith},\ and\ \citenamefont
  {Tiglio}}]{Canizares_2015}%
  \BibitemOpen
  \bibfield  {author} {\bibinfo {author} {\bibfnamefont {P.}~\bibnamefont
  {Canizares}}, \bibinfo {author} {\bibfnamefont {S.~E.}\ \bibnamefont
  {Field}}, \bibinfo {author} {\bibfnamefont {J.}~\bibnamefont {Gair}},
  \bibinfo {author} {\bibfnamefont {V.}~\bibnamefont {Raymond}}, \bibinfo
  {author} {\bibfnamefont {R.}~\bibnamefont {Smith}}, \ and\ \bibinfo {author}
  {\bibfnamefont {M.}~\bibnamefont {Tiglio}},\ }\href {\doibase
  10.1103/PhysRevLett.114.071104} {\bibfield  {journal} {\bibinfo  {journal}
  {Phys. Rev. Lett.}\ }\textbf {\bibinfo {volume} {114}},\ \bibinfo {pages}
  {071104} (\bibinfo {year} {2015})}\BibitemShut {NoStop}%
\bibitem [{\citenamefont {Smith}\ \emph {et~al.}(2016)\citenamefont {Smith},
  \citenamefont {Field}, \citenamefont {Blackburn}, \citenamefont {Haster},
  \citenamefont {P\"urrer}, \citenamefont {Raymond},\ and\ \citenamefont
  {Schmidt}}]{Smith_2016}%
  \BibitemOpen
  \bibfield  {author} {\bibinfo {author} {\bibfnamefont {R.}~\bibnamefont
  {Smith}}, \bibinfo {author} {\bibfnamefont {S.~E.}\ \bibnamefont {Field}},
  \bibinfo {author} {\bibfnamefont {K.}~\bibnamefont {Blackburn}}, \bibinfo
  {author} {\bibfnamefont {C.-J.}\ \bibnamefont {Haster}}, \bibinfo {author}
  {\bibfnamefont {M.}~\bibnamefont {P\"urrer}}, \bibinfo {author}
  {\bibfnamefont {V.}~\bibnamefont {Raymond}}, \ and\ \bibinfo {author}
  {\bibfnamefont {P.}~\bibnamefont {Schmidt}},\ }\href {\doibase
  10.1103/PhysRevD.94.044031} {\bibfield  {journal} {\bibinfo  {journal} {Phys.
  Rev. D}\ }\textbf {\bibinfo {volume} {94}},\ \bibinfo {pages} {044031}
  (\bibinfo {year} {2016})}\BibitemShut {NoStop}%
\bibitem [{\citenamefont {Cook}\ \emph {et~al.}(2006)\citenamefont {Cook},
  \citenamefont {Gelman},\ and\ \citenamefont
  {Rubin}}]{doi:10.1198/106186006X136976}%
  \BibitemOpen
  \bibfield  {author} {\bibinfo {author} {\bibfnamefont {S.~R.}\ \bibnamefont
  {Cook}}, \bibinfo {author} {\bibfnamefont {A.}~\bibnamefont {Gelman}}, \ and\
  \bibinfo {author} {\bibfnamefont {D.~B.}\ \bibnamefont {Rubin}},\ }\href
  {\doibase 10.1198/106186006X136976} {\bibfield  {journal} {\bibinfo
  {journal} {Journal of Computational and Graphical Statistics}\ }\textbf
  {\bibinfo {volume} {15}},\ \bibinfo {pages} {675} (\bibinfo {year} {2006})},\
  \Eprint {http://arxiv.org/abs/https://doi.org/10.1198/106186006X136976}
  {https://doi.org/10.1198/106186006X136976} \BibitemShut {NoStop}%
\bibitem [{\citenamefont {{Gair}}\ and\ \citenamefont
  {{Moore}}(2015)}]{gair2015quantifying}%
  \BibitemOpen
  \bibfield  {author} {\bibinfo {author} {\bibfnamefont {J.~R.}\ \bibnamefont
  {{Gair}}}\ and\ \bibinfo {author} {\bibfnamefont {C.~J.}\ \bibnamefont
  {{Moore}}},\ }\href {\doibase 10.1103/PhysRevD.91.124062} {\bibfield
  {journal} {\bibinfo  {journal} {\prd}\ }\textbf {\bibinfo {volume} {91}},\
  \bibinfo {eid} {124062} (\bibinfo {year} {2015})},\ \Eprint
  {http://arxiv.org/abs/1504.02767} {arXiv:1504.02767 [gr-qc]} \BibitemShut
  {NoStop}%
\bibitem [{\citenamefont {Berry}\ \emph {et~al.}(2015)\citenamefont {Berry},
  \citenamefont {Mandel}, \citenamefont {Middleton}, \citenamefont {Singer},
  \citenamefont {Urban}, \citenamefont {Vecchio} \emph {et~al.}}]{Berry_2015}%
  \BibitemOpen
  \bibfield  {author} {\bibinfo {author} {\bibfnamefont {C.~P.~L.}\
  \bibnamefont {Berry}}, \bibinfo {author} {\bibfnamefont {I.}~\bibnamefont
  {Mandel}}, \bibinfo {author} {\bibfnamefont {H.}~\bibnamefont {Middleton}},
  \bibinfo {author} {\bibfnamefont {L.~P.}\ \bibnamefont {Singer}}, \bibinfo
  {author} {\bibfnamefont {A.~L.}\ \bibnamefont {Urban}}, \bibinfo {author}
  {\bibfnamefont {A.}~\bibnamefont {Vecchio}},  \emph {et~al.},\ }\href
  {\doibase 10.1088/0004-637x/804/2/114} {\bibfield  {journal} {\bibinfo
  {journal} {The Astrophysical Journal}\ }\textbf {\bibinfo {volume} {804}},\
  \bibinfo {pages} {114} (\bibinfo {year} {2015})}\BibitemShut {NoStop}%
\bibitem [{\citenamefont {Sidery}\ \emph {et~al.}(2014)\citenamefont {Sidery},
  \citenamefont {Aylott}, \citenamefont {Christensen}, \citenamefont {Farr},
  \citenamefont {Farr}, \citenamefont {Feroz} \emph
  {et~al.}}]{PhysRevD.89.084060}%
  \BibitemOpen
  \bibfield  {author} {\bibinfo {author} {\bibfnamefont {T.}~\bibnamefont
  {Sidery}}, \bibinfo {author} {\bibfnamefont {B.}~\bibnamefont {Aylott}},
  \bibinfo {author} {\bibfnamefont {N.}~\bibnamefont {Christensen}}, \bibinfo
  {author} {\bibfnamefont {B.}~\bibnamefont {Farr}}, \bibinfo {author}
  {\bibfnamefont {W.}~\bibnamefont {Farr}}, \bibinfo {author} {\bibfnamefont
  {F.}~\bibnamefont {Feroz}},  \emph {et~al.},\ }\href {\doibase
  10.1103/PhysRevD.89.084060} {\bibfield  {journal} {\bibinfo  {journal} {Phys.
  Rev. D}\ }\textbf {\bibinfo {volume} {89}},\ \bibinfo {pages} {084060}
  (\bibinfo {year} {2014})}\BibitemShut {NoStop}%
\bibitem [{\citenamefont {Pankow}\ \emph {et~al.}(2015)\citenamefont {Pankow},
  \citenamefont {Brady}, \citenamefont {Ochsner},\ and\ \citenamefont
  {O'Shaughnessy}}]{PhysRevD.92.023002}%
  \BibitemOpen
  \bibfield  {author} {\bibinfo {author} {\bibfnamefont {C.}~\bibnamefont
  {Pankow}}, \bibinfo {author} {\bibfnamefont {P.}~\bibnamefont {Brady}},
  \bibinfo {author} {\bibfnamefont {E.}~\bibnamefont {Ochsner}}, \ and\
  \bibinfo {author} {\bibfnamefont {R.}~\bibnamefont {O'Shaughnessy}},\ }\href
  {\doibase 10.1103/PhysRevD.92.023002} {\bibfield  {journal} {\bibinfo
  {journal} {Phys. Rev. D}\ }\textbf {\bibinfo {volume} {92}},\ \bibinfo
  {pages} {023002} (\bibinfo {year} {2015})}\BibitemShut {NoStop}%
\bibitem [{\citenamefont {Hotokezaka}\ \emph {et~al.}(2019)\citenamefont
  {Hotokezaka}, \citenamefont {Nakar}, \citenamefont {Gottlieb}, \citenamefont
  {Nissanke}, \citenamefont {Masuda}, \citenamefont {Hallinan}, \citenamefont
  {Mooley},\ and\ \citenamefont {Deller}}]{Hotokezaka:2018dfi}%
  \BibitemOpen
  \bibfield  {author} {\bibinfo {author} {\bibfnamefont {K.}~\bibnamefont
  {Hotokezaka}}, \bibinfo {author} {\bibfnamefont {E.}~\bibnamefont {Nakar}},
  \bibinfo {author} {\bibfnamefont {O.}~\bibnamefont {Gottlieb}}, \bibinfo
  {author} {\bibfnamefont {S.}~\bibnamefont {Nissanke}}, \bibinfo {author}
  {\bibfnamefont {K.}~\bibnamefont {Masuda}}, \bibinfo {author} {\bibfnamefont
  {G.}~\bibnamefont {Hallinan}}, \bibinfo {author} {\bibfnamefont {K.~P.}\
  \bibnamefont {Mooley}}, \ and\ \bibinfo {author} {\bibfnamefont {A.~T.}\
  \bibnamefont {Deller}},\ }\href {\doibase 10.1038/s41550-019-0820-1}
  {\bibfield  {journal} {\bibinfo  {journal} {Nature Astron.}\ }\textbf
  {\bibinfo {volume} {3}},\ \bibinfo {pages} {940} (\bibinfo {year} {2019})},\
  \Eprint {http://arxiv.org/abs/1806.10596} {arXiv:1806.10596 [astro-ph.CO]}
  \BibitemShut {NoStop}%
\bibitem [{\citenamefont {{Mukherjee, Suvodip}}\ \emph
  {et~al.}(2021{\natexlab{b}})\citenamefont {{Mukherjee, Suvodip}},
  \citenamefont {{Lavaux, Guilhem}}, \citenamefont {{Bouchet, Fran\c{c}ois
  R.}}, \citenamefont {{Jasche, Jens}}, \citenamefont {{Wandelt, Benjamin D.}},
  \citenamefont {{Nissanke, Samaya}}, \citenamefont {{Leclercq, Florent}},\
  and\ \citenamefont {{Hotokezaka, Kenta}}}]{Mukherjee:2019qmm}%
  \BibitemOpen
  \bibfield  {author} {\bibinfo {author} {\bibnamefont {{Mukherjee, Suvodip}}},
  \bibinfo {author} {\bibnamefont {{Lavaux, Guilhem}}}, \bibinfo {author}
  {\bibnamefont {{Bouchet, Fran\c{c}ois R.}}}, \bibinfo {author} {\bibnamefont
  {{Jasche, Jens}}}, \bibinfo {author} {\bibnamefont {{Wandelt, Benjamin D.}}},
  \bibinfo {author} {\bibnamefont {{Nissanke, Samaya}}}, \bibinfo {author}
  {\bibnamefont {{Leclercq, Florent}}}, \ and\ \bibinfo {author} {\bibnamefont
  {{Hotokezaka, Kenta}}},\ }\href {\doibase 10.1051/0004-6361/201936724}
  {\bibfield  {journal} {\bibinfo  {journal} {Astron. Astrophys.}\ }\textbf
  {\bibinfo {volume} {646}},\ \bibinfo {pages} {A65} (\bibinfo {year}
  {2021}{\natexlab{b}})},\ \Eprint {http://arxiv.org/abs/1909.08627}
  {arXiv:1909.08627 [astro-ph.CO]} \BibitemShut {NoStop}%
\bibitem [{\citenamefont {Fishbach}\ \emph {et~al.}(2019)\citenamefont
  {Fishbach}, \citenamefont {Gray}, \citenamefont {Hernandez}, \citenamefont
  {Qi}, \citenamefont {Sur}, \citenamefont {Acernese} \emph
  {et~al.}}]{Fishbach_2019}%
  \BibitemOpen
  \bibfield  {author} {\bibinfo {author} {\bibfnamefont {M.}~\bibnamefont
  {Fishbach}}, \bibinfo {author} {\bibfnamefont {R.}~\bibnamefont {Gray}},
  \bibinfo {author} {\bibfnamefont {I.~M.}\ \bibnamefont {Hernandez}}, \bibinfo
  {author} {\bibfnamefont {H.}~\bibnamefont {Qi}}, \bibinfo {author}
  {\bibfnamefont {A.}~\bibnamefont {Sur}}, \bibinfo {author} {\bibfnamefont
  {F.}~\bibnamefont {Acernese}},  \emph {et~al.},\ }\href {\doibase
  10.3847/2041-8213/aaf96e} {\bibfield  {journal} {\bibinfo  {journal} {The
  Astrophysical Journal}\ }\textbf {\bibinfo {volume} {871}},\ \bibinfo {pages}
  {L13} (\bibinfo {year} {2019})}\BibitemShut {NoStop}%
\bibitem [{\citenamefont {Soares-Santos}\ \emph {et~al.}(2019)\citenamefont
  {Soares-Santos}, \citenamefont {Palmese}, \citenamefont {Hartley},
  \citenamefont {Annis}, \citenamefont {Garcia-Bellido}, \citenamefont {Lahav}
  \emph {et~al.}}]{Soares_Santos_2019}%
  \BibitemOpen
  \bibfield  {author} {\bibinfo {author} {\bibfnamefont {M.}~\bibnamefont
  {Soares-Santos}}, \bibinfo {author} {\bibfnamefont {A.}~\bibnamefont
  {Palmese}}, \bibinfo {author} {\bibfnamefont {W.}~\bibnamefont {Hartley}},
  \bibinfo {author} {\bibfnamefont {J.}~\bibnamefont {Annis}}, \bibinfo
  {author} {\bibfnamefont {J.}~\bibnamefont {Garcia-Bellido}}, \bibinfo
  {author} {\bibfnamefont {O.}~\bibnamefont {Lahav}},  \emph {et~al.},\ }\href
  {\doibase 10.3847/2041-8213/ab14f1} {\bibfield  {journal} {\bibinfo
  {journal} {The Astrophysical Journal}\ }\textbf {\bibinfo {volume} {876}},\
  \bibinfo {pages} {L7} (\bibinfo {year} {2019})}\BibitemShut {NoStop}%
\bibitem [{\citenamefont {Gray}\ \emph {et~al.}(2020)\citenamefont {Gray},
  \citenamefont {Hernandez}, \citenamefont {Qi}, \citenamefont {Sur},
  \citenamefont {Brady}, \citenamefont {Chen} \emph {et~al.}}]{Gray_2020}%
  \BibitemOpen
  \bibfield  {author} {\bibinfo {author} {\bibfnamefont {R.}~\bibnamefont
  {Gray}}, \bibinfo {author} {\bibfnamefont {I.~M.~n.}\ \bibnamefont
  {Hernandez}}, \bibinfo {author} {\bibfnamefont {H.}~\bibnamefont {Qi}},
  \bibinfo {author} {\bibfnamefont {A.}~\bibnamefont {Sur}}, \bibinfo {author}
  {\bibfnamefont {P.~R.}\ \bibnamefont {Brady}}, \bibinfo {author}
  {\bibfnamefont {H.-Y.}\ \bibnamefont {Chen}},  \emph {et~al.},\ }\href
  {\doibase 10.1103/PhysRevD.101.122001} {\bibfield  {journal} {\bibinfo
  {journal} {Phys. Rev. D}\ }\textbf {\bibinfo {volume} {101}},\ \bibinfo
  {pages} {122001} (\bibinfo {year} {2020})}\BibitemShut {NoStop}%
\bibitem [{\citenamefont {Finke}\ \emph {et~al.}(2021)\citenamefont {Finke},
  \citenamefont {Foffa}, \citenamefont {Iacovelli}, \citenamefont {Maggiore},\
  and\ \citenamefont {Mancarella}}]{Finke_2021}%
  \BibitemOpen
  \bibfield  {author} {\bibinfo {author} {\bibfnamefont {A.}~\bibnamefont
  {Finke}}, \bibinfo {author} {\bibfnamefont {S.}~\bibnamefont {Foffa}},
  \bibinfo {author} {\bibfnamefont {F.}~\bibnamefont {Iacovelli}}, \bibinfo
  {author} {\bibfnamefont {M.}~\bibnamefont {Maggiore}}, \ and\ \bibinfo
  {author} {\bibfnamefont {M.}~\bibnamefont {Mancarella}},\ }\href {\doibase
  10.1088/1475-7516/2021/08/026} {\bibfield  {journal} {\bibinfo  {journal}
  {Journal of Cosmology and Astroparticle Physics}\ }\textbf {\bibinfo {volume}
  {2021}},\ \bibinfo {pages} {026} (\bibinfo {year} {2021})}\BibitemShut
  {NoStop}%
\bibitem [{\citenamefont {Abbott}\ \emph
  {et~al.}(2021{\natexlab{e}})\citenamefont {Abbott} \emph
  {et~al.}}]{LVK:2021bbr}%
  \BibitemOpen
  \bibfield  {author} {\bibinfo {author} {\bibfnamefont {R.}~\bibnamefont
  {Abbott}} \emph {et~al.} (\bibinfo {collaboration} {LIGO Scientific, Virgo,
  KAGRA}),\ }\href@noop {} {\  (\bibinfo {year} {2021}{\natexlab{e}})},\
  \Eprint {http://arxiv.org/abs/2111.03604} {arXiv:2111.03604 [astro-ph.CO]}
  \BibitemShut {NoStop}%
\bibitem [{\citenamefont {Punturo}\ \emph {et~al.}(2010)\citenamefont
  {Punturo}, \citenamefont {Abernathy}, \citenamefont {Acernese}, \citenamefont
  {Allen}, \citenamefont {Andersson}, \citenamefont {Arun} \emph
  {et~al.}}]{Punturo:2010zz}%
  \BibitemOpen
  \bibfield  {author} {\bibinfo {author} {\bibfnamefont {M.}~\bibnamefont
  {Punturo}}, \bibinfo {author} {\bibfnamefont {M.}~\bibnamefont {Abernathy}},
  \bibinfo {author} {\bibfnamefont {F.}~\bibnamefont {Acernese}}, \bibinfo
  {author} {\bibfnamefont {B.}~\bibnamefont {Allen}}, \bibinfo {author}
  {\bibfnamefont {N.}~\bibnamefont {Andersson}}, \bibinfo {author}
  {\bibfnamefont {K.}~\bibnamefont {Arun}},  \emph {et~al.},\ }\href {\doibase
  10.1088/0264-9381/27/19/194002} {\bibfield  {journal} {\bibinfo  {journal}
  {Class. Quant. Grav.}\ }\textbf {\bibinfo {volume} {27}},\ \bibinfo {pages}
  {194002} (\bibinfo {year} {2010})}\BibitemShut {NoStop}%
\bibitem [{\citenamefont {Maggiore}\ \emph {et~al.}(2020)\citenamefont
  {Maggiore}, \citenamefont {Broeck}, \citenamefont {Bartolo}, \citenamefont
  {Belgacem}, \citenamefont {Bertacca}, \citenamefont {Bizouard} \emph
  {et~al.}}]{Maggiore:2019uih}%
  \BibitemOpen
  \bibfield  {author} {\bibinfo {author} {\bibfnamefont {M.}~\bibnamefont
  {Maggiore}}, \bibinfo {author} {\bibfnamefont {C.~V.~D.}\ \bibnamefont
  {Broeck}}, \bibinfo {author} {\bibfnamefont {N.}~\bibnamefont {Bartolo}},
  \bibinfo {author} {\bibfnamefont {E.}~\bibnamefont {Belgacem}}, \bibinfo
  {author} {\bibfnamefont {D.}~\bibnamefont {Bertacca}}, \bibinfo {author}
  {\bibfnamefont {M.~A.}\ \bibnamefont {Bizouard}},  \emph {et~al.},\ }\href
  {\doibase 10.1088/1475-7516/2020/03/050} {\bibfield  {journal} {\bibinfo
  {journal} {JCAP}\ }\textbf {\bibinfo {volume} {03}},\ \bibinfo {pages} {050}
  (\bibinfo {year} {2020})},\ \Eprint {http://arxiv.org/abs/1912.02622}
  {arXiv:1912.02622 [astro-ph.CO]} \BibitemShut {NoStop}%
\bibitem [{\citenamefont {Abbott}\ \emph
  {et~al.}(2017{\natexlab{c}})\citenamefont {Abbott} \emph
  {et~al.}}]{LIGOScientific:2016wof}%
  \BibitemOpen
  \bibfield  {author} {\bibinfo {author} {\bibfnamefont {B.~P.}\ \bibnamefont
  {Abbott}} \emph {et~al.} (\bibinfo {collaboration} {LIGO Scientific}),\
  }\href {\doibase 10.1088/1361-6382/aa51f4} {\bibfield  {journal} {\bibinfo
  {journal} {Class. Quant. Grav.}\ }\textbf {\bibinfo {volume} {34}},\ \bibinfo
  {pages} {044001} (\bibinfo {year} {2017}{\natexlab{c}})},\ \Eprint
  {http://arxiv.org/abs/1607.08697} {arXiv:1607.08697 [astro-ph.IM]}
  \BibitemShut {NoStop}%
\bibitem [{\citenamefont {Reitze}\ \emph {et~al.}(2019)\citenamefont {Reitze},
  \citenamefont {Adhikari}, \citenamefont {Ballmer}, \citenamefont {Barish},
  \citenamefont {Barsotti}, \citenamefont {Billingsley} \emph
  {et~al.}}]{Reitze:2019iox}%
  \BibitemOpen
  \bibfield  {author} {\bibinfo {author} {\bibfnamefont {D.}~\bibnamefont
  {Reitze}}, \bibinfo {author} {\bibfnamefont {R.~X.}\ \bibnamefont
  {Adhikari}}, \bibinfo {author} {\bibfnamefont {S.}~\bibnamefont {Ballmer}},
  \bibinfo {author} {\bibfnamefont {B.}~\bibnamefont {Barish}}, \bibinfo
  {author} {\bibfnamefont {L.}~\bibnamefont {Barsotti}}, \bibinfo {author}
  {\bibfnamefont {G.}~\bibnamefont {Billingsley}},  \emph {et~al.},\
  }\href@noop {} {\bibfield  {journal} {\bibinfo  {journal} {Bull. Am. Astron.
  Soc.}\ }\textbf {\bibinfo {volume} {51}},\ \bibinfo {pages} {035} (\bibinfo
  {year} {2019})},\ \Eprint {http://arxiv.org/abs/1907.04833} {arXiv:1907.04833
  [astro-ph.IM]} \BibitemShut {NoStop}%
\end{thebibliography}%

\end{document}